\documentclass[aps,prl,reprint,groupedaddress]{revtex4-1}

\usepackage{graphicx}
\usepackage{amsmath}
\usepackage{multirow}

\newlength{\digit}
\newcommand{\dd}{{\hspace*{\digit}}} 
\newcommand{\tc}[1]{\multicolumn{1}{c}{{#1}}}
\newcommand\tstrut{\rule{0pt}{2.5ex}}

\newcommand{\NF}{{$N_\mathrm{frag}\/$ }}

\def\apj{ApJ}%
\def\apjl{ApJ}%
\def\aap{A\&A}%
\def\icarus{Icarus}%
\def\mnras{MNRAS}%

\begin{document}


\title{Planetary systems and the Formation of Habitable Planets}




\author{Rudolf Dvorak}
\email[]{rudolf.dvorak@univie.ac.at}
\author{Thomas I.\ Maindl}
\author{Christoph Burger}
\affiliation{Universit\"atssternwarte Wien, T\"urkenschanzstr. 17, 1180, Wien, Austria}

\author{Christoph Sch\"afer}
\affiliation{Institut f\"ur Astronomie und Astrophysik, Eberhard Karls Universit\"at T\"ubingen, Auf der Morgenstelle 10, 72076 T\"ubingen, Germany}

\author{Roland Speith}
\affiliation{Physikalisches Institut, Eberhard Karls Universit\"at T\"ubingen, Auf der Morgenstelle 14, 72076 T\"ubingen, Germany}



\date{\today}

\begin{abstract}
As part of a big
national scientific network 'Pathways to Habitable Worlds' the formation of
planets and the delivery of 
water onto these planets is a key question because water is essential for the 
development of life on a planet. In the first part of the paper we summarize the state of the art of planet formation -- which is still under debate in the astronomical
community -- before
we show our results on the formation of planets. The outcome of our numerical simulations depends a lot on the choice of the initial distribution of planetesimals and 
planetary embryos after the disappearance of gas in the protoplanetary disk. 
In addition we take into account that some of these planetesimals of sizes in the order
of the mass of the Moon already contained water; the quantity depends on the
distance to the Sun - too close and the bodies are dry, but starting from a distance of about 2 AU they can contain substantial amounts of water.
Our assumption is that the gas giants and the
terrestrial planets are already formed when we check the collisions of the  
small bodies containing water (in the order of a few percent) 
with the terrestrial planets. We thus are able to give an estimate 
of the respective contribution to the actual water content (of some Earth-oceans) 
in the mantle, in the crust and on the surface of Earth.
In the second part we discuss how the formation of larger bodies after 
a collision may happen in detail, because the outcome depends on different
parameters like the collision velocity, the angle of the impact and the materials
involved. The detailed results were accomplished with the aid of SPH (Smoothed
Particle Hydrodynamics) simulations, a sophisticated numerical method to simulate 
these events. We briefly describe this method and show different scenarios
with respect to the formed bodies, possible fragmentation and the water
content before and after the collision.\\
In an appendix we discuss the detection methods for extrasolar planets around
other stars for which the current number of known ones is already close to 2000.

\end{abstract}



\maketitle



\section{state of the art}

The formation of terrestrial planets is an outstanding problem in astronomy
which has -- up to now -- not been solved satisfactory. Especially difficult is
to form a low-mass Mars at about 1.5 AU and a very dense Mercury inside the
orbit of Venus. All of the theories in this respect assume that the gas
giants in the outer Solar System were formed first from a
gas dominated protoplanetary disk in relatively short time \citep{2007astro.ph..1485A} and that the inner planets were formed later.
Globally one can say that the outcome of every simulation
is highly dependent on the initial conditions for the distribution of mass in
the early Solar System. The Mars problem has led to different approaches
concerning the scenarios for the growth of planetesimals to planetary 
embryos to protoplanets and finally to planets. Because of the small mass -
1/10 of Earth's mass for Mars (and Mercury) - they are regarded as not
fully developed planets, but rather as protoplanets which did not accumulate mass in
later stages of the early history of the Solar System. While for Earth (and
also for Venus) a formation time of about 100 million years seems appropriate,
Mars should have formed within 10 million years only. One recent theory is 
the so-called GrandTack model \citep{2011Natur.475..206W}:

\par
\begingroup
\leftskip5mm
A disk (up to 3 AU) of relatively dry planetesimals in low-eccentricity orbits resides inside an already fully formed Jupiter at 
about 3.5 AU and Saturn at about 5 AU still accumulating material. The
inward migration of the gas giants caused by the still present gas 
drag truncates the original disk to an outer edge of 1 AU. When Jupiter 
was at 1.5 AU and Saturn (now fully formed) was caught in a 2:3 mean motion
resonance at about 2 AU the migration was reversed. Without a deeper
explanation in the original paper it was stated that ' ... according to our
results Jupiter tacked at 1.5 AU and then migrated outward owing to the
presence of Saturn' \citep{2011Natur.475..206W}. During this stage of migration
outward, outer wet material was scattered to the
region inside Jupiter's orbit, so that finally the distribution of the 
asteroid main belt, composed of a dry inner region (S-type asteroids), and a wet 
outer region (C-type asteroids), as we see it today, was reached, with the 
gas giants more or less in their present positions. Now the formation of the terrestrial planets to their final masses and compositions took place in this inner region, which already started during the outward migration of
the gas giants.
\par
\endgroup
\vspace{2.5mm}
Prior to that interesting but very special scenario discussed before, where
many assumptions needed to be made, a simple simulation \citep{2009ApJ...703.1131H},
starting with 400 equally massive bodies (0.005 Earth-masses each), randomly distributed 
in an annulus between 0.7 and 1 AU  with low eccentricities and inclinations,
has shown a surprising result.
The distribution of the terrestrial planets after long term integrations turned out to be the one which we know from our present-day Solar System:
a small Mercury inside, and a small Mars outside the orbits of the 'twin planets' Earth and Venus close to their present positions. 

A recently published theory \citep{2014ApJ...782...31I} uses these results and
explains the existence of a low-mass Mars by straight forward integrations
with a special assumption for the distribution of about 150 planetary embryos
and about 1000 planetesimals inside a disk extending from 0.5 AU to 4 AU 
in the presence of
Jupiter and Saturn in their current positions. Their major point is that they
assume a distribution of planetesimals and embryos being much lower
around 1.5 AU, where Mars can be found today. According to an article by
\citet{2008ApJ...674L.105J} this local minimum can be explained by the 
evolution of the surface
density distribution of the solar nebula during the first million years. But as
pointed out by \citet{2009Icar..203..644R} and \citet{2014prpl.conf..595R} this dip 
is too narrow to cut off Mars' accretion. Given the broadness of the lower-density gap used by \citet{2014ApJ...782...31I} the results shown there should be
taken with caution.    
   
It should be emphasized that for the last 20 years the interest in studies about the
formation of planets and especially terrestrial planets with respect to their
habitability is
growing quasi exponentially and the literature of only the last years is huge.
There are some summarizing review articles on this topic which more or less
discuss the latest progress, but all of these papers are very much directed
towards the authors' own research and leave many questions out. Just a few of them should
be mentioned: \citet{2004ApJ...613L.157A}, \citet{2009Icar..203..644R}, \citet{2010ApJ...719L..45M}, \citet{2012ApJ...745..143A}, \citet{2012AREPS..40..251M}, \citet{2014prpl.conf..595R}, \citet{2012AREPS..40..597L} and \citet{2014prpl.conf..883G}.
   
\section{Our Methods}

One main constraint in all the cited simulations is the number of integrated
planetesimals and planetary embryos which was always limited according to the
integration times on single CPUs. Although some interesting codes are
available, where up to about 1000 bodies can be integrated in the Newtonian
framework (e.g. \citet{1999MNRAS.304..793C}), these simulations are not precise in a
very important fact: the collisions of two bodies which merge are not
precisely computed, which causes a major drawback. In this connection we were
and are still working into different directions: first of all we avoid
simplifications for collisions and the merging of bodies insofar as that we compute
the close encounters up to the moment when the two bodies involved are actually 
touching; secondly we extended the programs to be able to integrate several
thousands of interacting bodies by writing codes running on graphic cards.
In the following we will concentrate on the first point.

\subsection{The Encounters and the Collisions}
As mentioned one crucial step in the process of forming planets are the
collisions of the early Solar System bodies and their possible
merging. Checking such an encounter in detail one can see that as well as the
impact velocity, the impact angle is a different one when we assume a
collision happening already at a larger distance than the real physical one as depicted in Fig.1
and Fig.2.

\begin{figure}
\includegraphics[width=2.7in,angle=0]{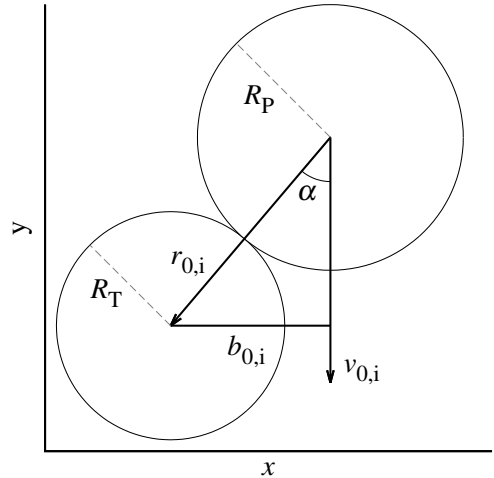}
\caption{Two colliding bodies at the moment of the first encounter: 
the two important parameters
are the impact velocity and the impact angle $\alpha$ besides their radii
$R_P$ and $R_T$. }
\label{collscheme}                                                              
\end{figure}

\begin{figure}     
\includegraphics[width=3.4in,height=3.4in,angle=270]{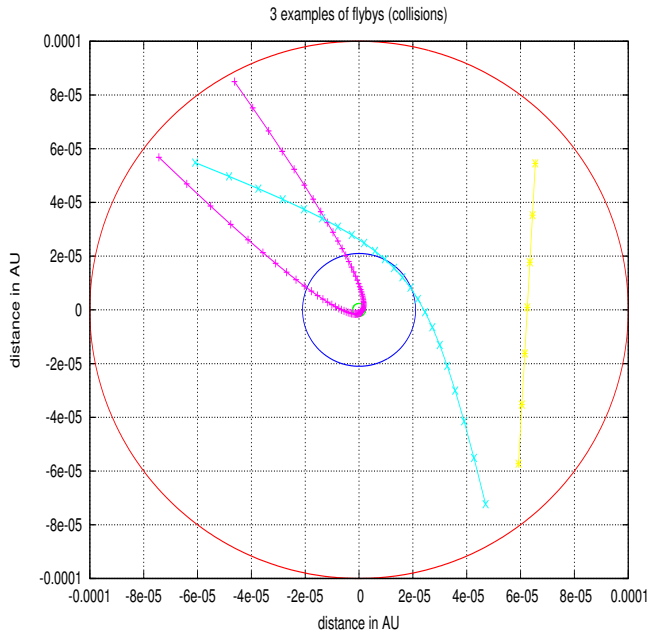}       
\caption{Planetocentric encounters with three different velocities and impact
  angles when entering a distance of 0.001 AU (red circle): a high 
velocity encounter (magenta), a moderate velocity encounter (blue) and a relatively far away fly by (yellow) are depicted (for more details see the text).}
\label{3ex}                                                              
\end{figure}

\begin{figure}    
\includegraphics[width=2.3in,angle=270]{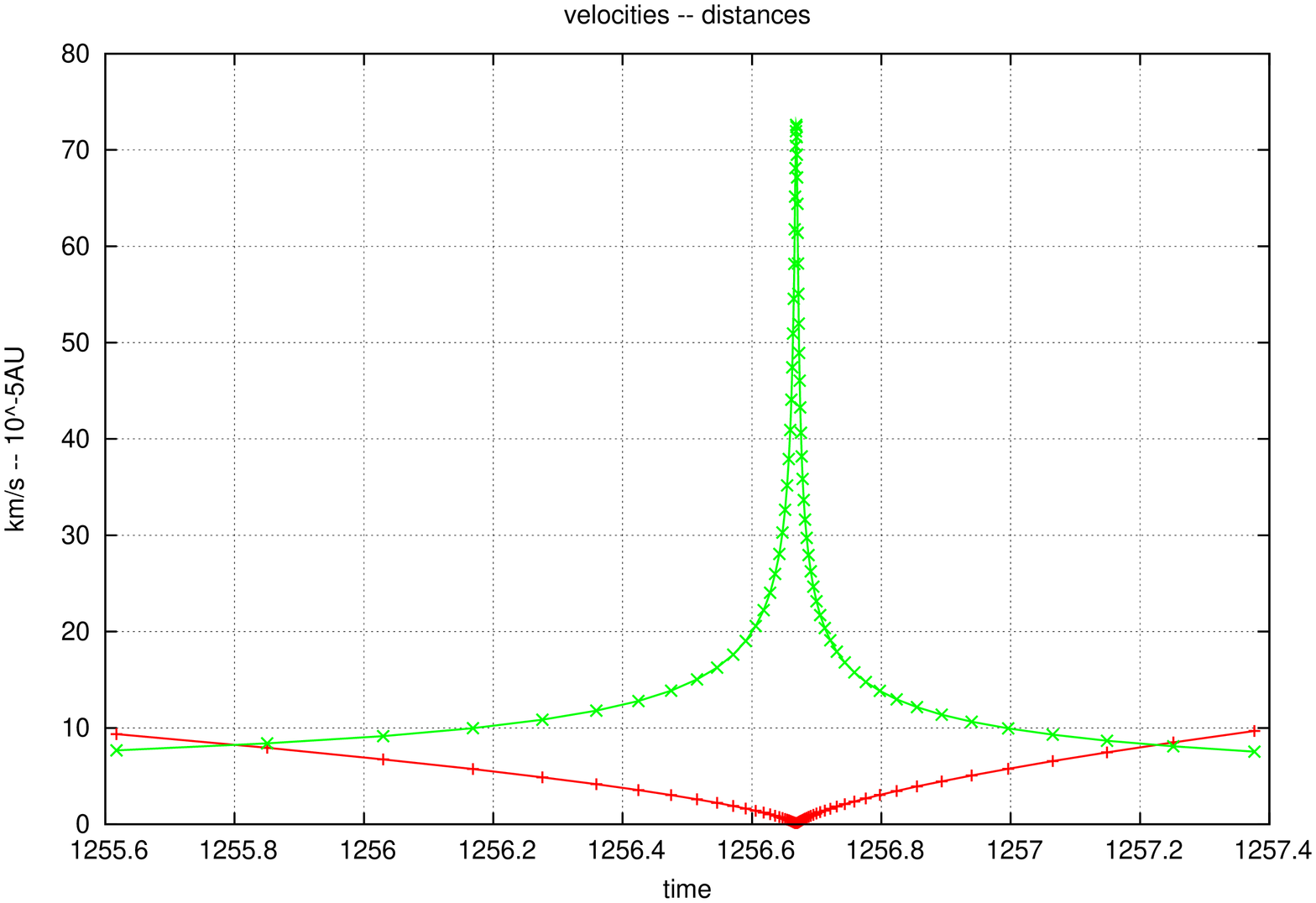} 
\includegraphics[width=2.3in,angle=270]{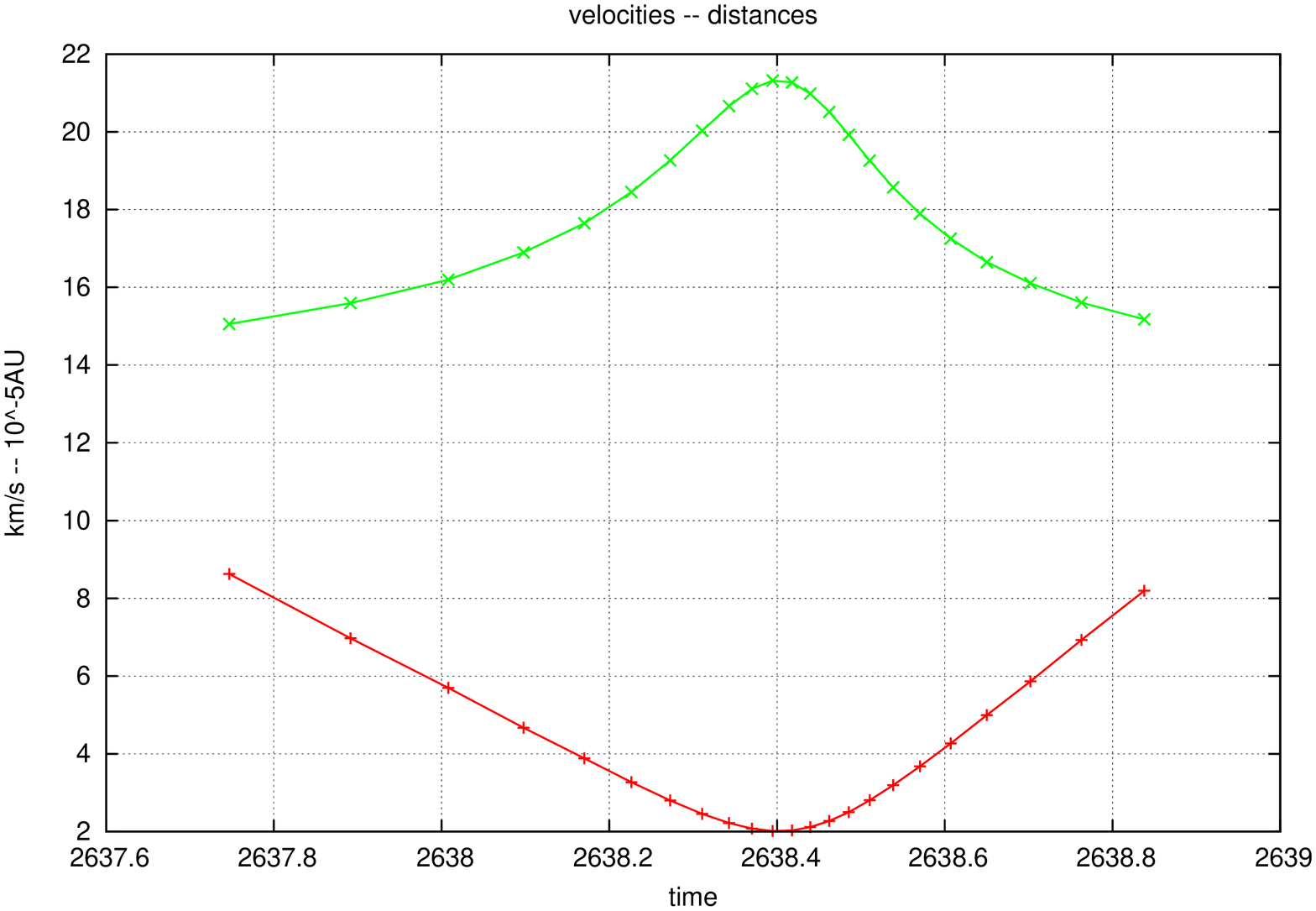}                              
\caption{Planetocentric distances (red) and collision velocities (green)
plotted versus time. A high velocity encounter (upper graph) and a low 
velocity encounter (lower graph) are shown (for more details see text).}
\label{disvel}                                                              
\end{figure}

The increasing velocity visible in Fig.\ref{disvel} (green line)
leads to a sharp cusp at the closest distance (red line). In fact the
collision between two extended bodies (e.g. two Ceres-sized bodies) 
and the following merging happens well
before the moment of entering the innermost small circle (green, $r_{0,i}$ for
two bodies of the mass of 1/10 of Ceres) in
Fig.\ref{3ex} and depends on the diameters of the two colliding bodies.
The second orbit in this figure seems to be tangential to the blue circle
representing the collision distance ($r_{0,i}$, depicted in
Fig.\ref{collscheme} for two arbitrary objects).

\begin{figure*}
\includegraphics[width=5.9in,angle=0]{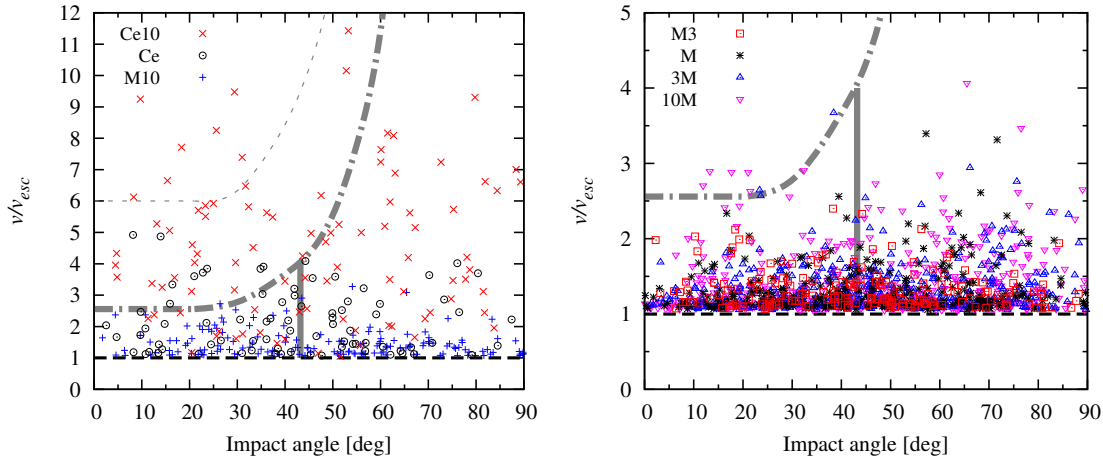}
\caption{Collision velocities (normalized by the surface escape velocities)
  versus impact angle for the collision of differently sized bodies from
  $m_{Ceres}/10$ to $10\ m_{Moon}$ (see detailed explanation in table 1 and in
  the text).}
\label{disruption}   
\end{figure*}

It has to be mentioned that different velocities and collision angles
lead to different outcomes, e.g. disruption, partial accretion and smooth merging, depicted in Fig.\ref{disruption}. 
Seven different masses were chosen to study the collision velocities, ranging from 1/10 of Ceres to 10 times the mass of the Moon (for explanation see table 1);
note that very high velocities only occur in the 1/10 Ceres scenario. The lines 
in the illustration refer to the boundaries of different collision outcomes as 
given in the paper by \citet{leiste12}: erosion of the 
target above the dot-dashed curve, below it partial
accretion to the left and hit-and-run scenarios to the right of the thick
vertical line. 
Finally a thin region of perfect merging can be found just above the lowermost dashed line, while below it (where $v/v_\mathrm{esc} < 1$) no collision is possible at all.
A more qualitative illustration of this behaviour can be found in Fig.\ref{fig:collmap}.

\begin{table}
\begin{tabular}{lllclrr}
\hline
\hline
body & Scenario & $m\; (\mathrm{M}_\odot)$ & $m\; (\mathrm{kg})$ & 
$R_\mathrm{imp}$ & $r_\mathrm{Hill}$ & $v_\mathrm{esc}\;(\mathrm{m}/\mathrm{s})$\\
\hline
$m_{Ceres*0.1}$ & Ce10 & $5\cdot 10^{-11}$ & $9.95\cdot 10^{19}$ & 0.3918 &  38 &  247\\
Ceres & Ce   & $5\cdot 10^{-10}$ & $9.95\cdot 10^{20}$ & 0.8440 &  82 &  532\\
$m_{Moon*0.1}$ & M10  & $3\cdot 10^{-9}$  & $5.97\cdot 10^{21}$ & 1.534  & 150 &  967\\
$m_{Moon*0.3}$ & M3   & $1\cdot 10^{-8}$  & $1.99\cdot 10^{22}$ & 2.291  & 223 & 1444\\
Moon & M    & $3\cdot 10^{-8}$  & $5.97\cdot 10^{22}$ & 3.304  & 322 & 2083\\
$m_{3*Moon}$ & 3M   & $1\cdot 10^{-7}$  & $1.99\cdot 10^{23}$ & 4.936  & 481 & 3112\\
$m_{10*Moon}$ & 10M  & $3\cdot 10^{-7}$  & $5.97\cdot 10^{23}$ & 7.119  & 694 & 4488\\
\hline
\end{tabular}
\caption{N-body simulation scenarios -- $m\/$ is the planetesimal mass,
  $R_\mathrm{imp}$ (in $10^6\,m$) denotes the mutual distance of the planetesimals' barycenters upon impact for $r_\mathrm{Hill}$ the Hill radius at 1\,AU and zero eccentricity, and $v_\mathrm
{esc}$ the target's surface escape velocity (see text); All scenarios include 750 planetesimals and are integrated for 1\,Myr.}
\label{t:nbodyscenarios}
\end{table}

It is evident from Fig.\ref{disruption} that the assumption of a simple inelastic merging for all collisions is not true and an oversimplification. 
Except for some notable exceptions (e.g. \citet{2012ApJ...744..137G}) nobody has yet accounted for this rather diverse range of collision outcomes.
We will concentrate on this issue in the subsequent sections when we discuss
results of SPH simulations of collisions.
Commonly the scenario of complete merging is used as an approach to the problem; in the next
section we show results of our own computations where we made 
these same ad hoc assumptions.

\section{Results of new computations}
As already mentioned most current scenarios of planet formation are lacking two
major points: the 'correct' initial distribution of the planetesimals/planetary
embryos and realistic collision outcomes. Whereas the
second point will be taken into account in future models only, the first 
point was considered by using the outcome of the Grand Tack model: we
distributed 100 dry planetesimals and planetary embryos randomly (with small
inclinations and eccentricities) in the range of 0.5 AU $<$ a $<$ 2.2 AU. 
Jupiter and Saturn were in their current positions, while Uranus and Neptune
were not present in the simulated debris disk. Whenever
two small bodies suffer from a collision the newly formed body replaces them
and a new smaller body (with low water content) is added from the outer
'wet' belt of planetesimals. This -- strictly speaking -- somehow artificial 
procedure was adopted due to numerical issues and aims to simulate the water transport to the inner 
dry region. The reason is to keep the number of bodies limited to 100, which is in 
our numerical approach, the Lie-series technique 
\citep{1984A&A...132..203H}, a number of bodies that allows to stay in 
a reasonable amount of computing time. Without going into the details of the numerical integration 
procedure it should be noted that the automatic step-size control makes 
the computation more accurate compared to other
methods \citep{2010LNP...790..431E} (especially concerning close encounters).

\begin{figure*}
\includegraphics[width=3.5in,angle=270]{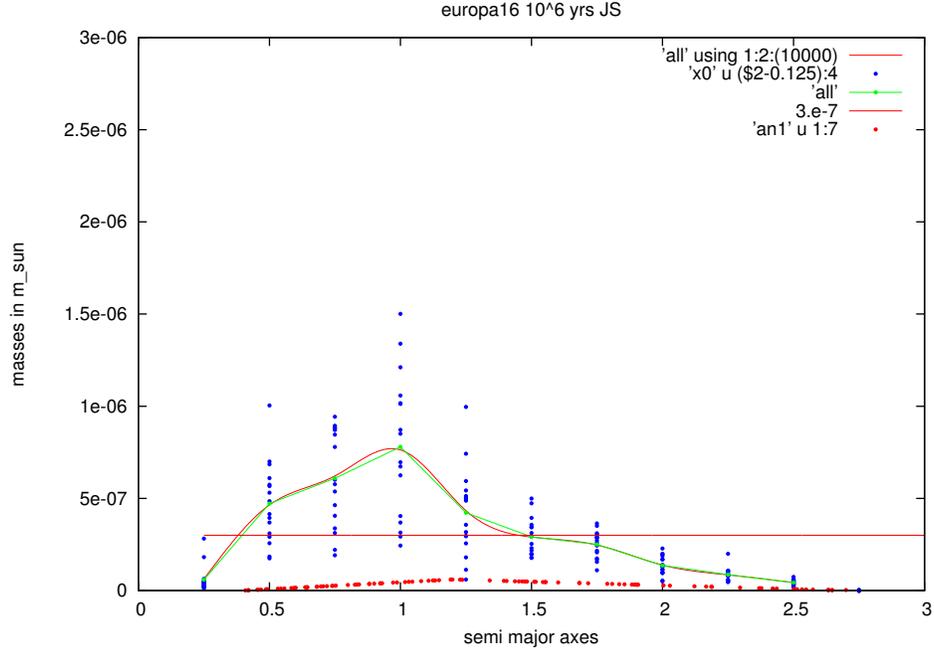}
\caption{Synopsis of simulations after 1.5 Myrs only with Jupiter 
for 16 different
initial conditions distributed according to the outcome of the GT scenario (red
points). The plot shows the merging planetesimals' semimajor-axes versus the mass in solar masses. A peak at 1 AU, corresponding to the
position of Earth, is clearly visible.}
\label{res16}
\end{figure*}

\begin{figure*}
\includegraphics[width=3.5in,angle=270]{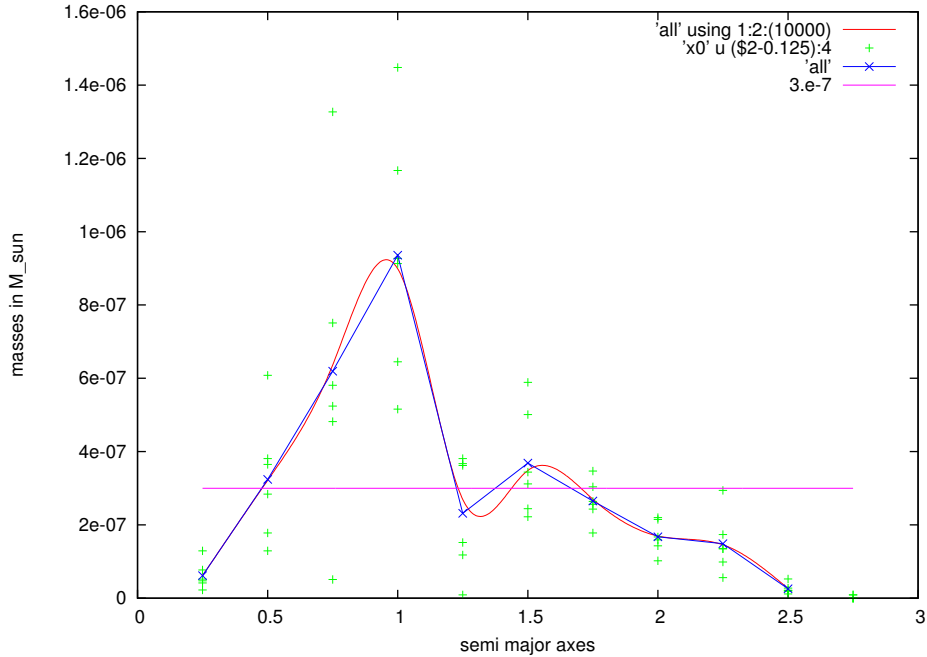}
\caption{Synopsis of simulations after 1.5 Myrs with Jupiter and Saturn for 6 different
  initial conditions distributed according to the outcome of the GT scenario. The plot shows the merging planetesimals' semimajor-axes versus the mass in solar masses. Two peaks are visible, one at 1 AU and another one at about 1.5 AU. These two peaks correspond approximately to the positions of Earth 
  and Mars. The magenta line marks a mass of 10 times the Moon's.}
\label{res6}
\end{figure*}

In the respective Fig.\ref{res16} we show the summarized results of 16 
independent simulations with different initial conditions with Jupiter 
 as sole perturbing planet of the cloud of planetesimals (embryos). The small
 red points mark the initial conditions for the 100 primary bodies, the dots
 represent the binned outcome after 1.5 Myrs. Two different interpolations 
show the peak at a=1 AU which is the position of Earth; the body with 
the largest mass has approximately half the mass of present-day Earth.

Another run was performed with only six independent simulations,
shown in Fig.\ref{res6}, but with Jupiter and Saturn as
perturbers of the belt of bodies inside Jupiter. Here one can observe two 
accumulation maxima, representing protoplanets, due to repeated merging processes: one at a=1 AU and
another one with a=1.5 AU. The second peak with a considerable lower mass
(about 1/10 of Earth's mass) is exactly at the location of Mars.

\begin{figure}
\includegraphics[width=3.75in,angle=0]{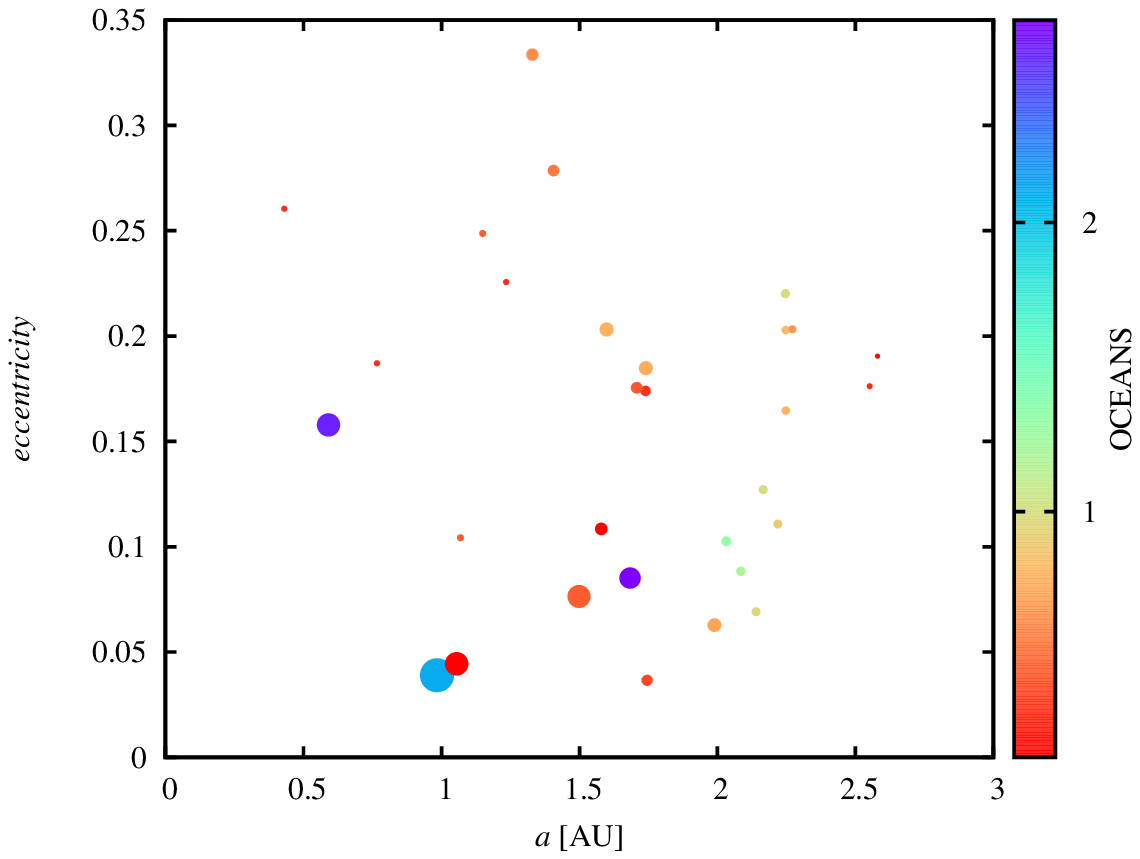}

\includegraphics[width=3.75in,angle=0]{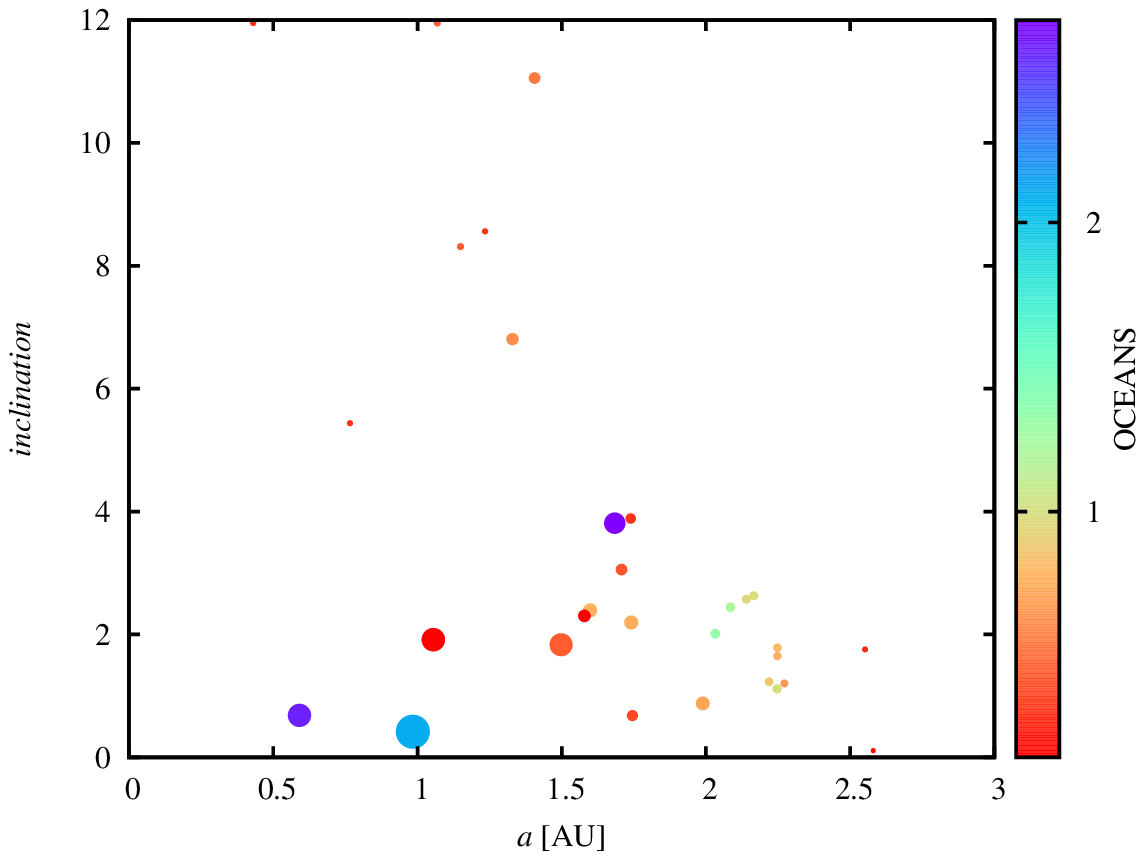} 
\caption{Results of planet formation after 1.5 Myrs: semimajor axis versus
  eccentricity (upper picture) and versus inclination (lower picture) respectively. The size of the points is proportional to their mass; the
  largest body formed at 1 AU and has about 30 times the mass of the Moon. The color code indicates
  the mass of water on the formed planets in Earth-oceans (for detailed
  explanation see text).}
\label{g21}
\end{figure}

One representative example out of the 22 different simulations is depicted in 
Fig.\ref{g21}. After 1.5 million years a planet of ca. 1/2 of Earth's mass is formed in a distance of almost 1 AU (planet
{\bf A}). This
planet has a low eccentricity and a very small inclination. Four more smaller
planets (embryos with a mass around 10 times the Moon's mass) are formed: 

\begin{enumerate}
\item Planet {\bf B}: at 0.59 AU, a Mercury like object with a relatively large eccentricity but a small inclination
\item Planet {\bf C}: at 0.98 AU, an object very close to planet {\bf A} with almost the same
  eccentricity but a slightly larger inclination
\item Planet {\bf D}: at 1.5 AU another embryo with eccentricity and inclination 
comparable to planet {\bf C} in the distance of Mars in our Solar System
\item Planet {\bf E}: at 1.7 AU, an object with an eccentricity comparable to {\bf D} but with a
  large inclination.
\end{enumerate}

The planets {\bf A} and {\bf C} were separately integrated for another 10 Myrs
and it turned out that after 6 Myrs they merge to a bigger body at approximately 1
AU. This shows that to form Earth-sized  planets the time scale is at least in
the order of tenth of Myrs. 

\subsection{The Watertransport}

\begin{figure}
\includegraphics[width=2.2in,angle=0]{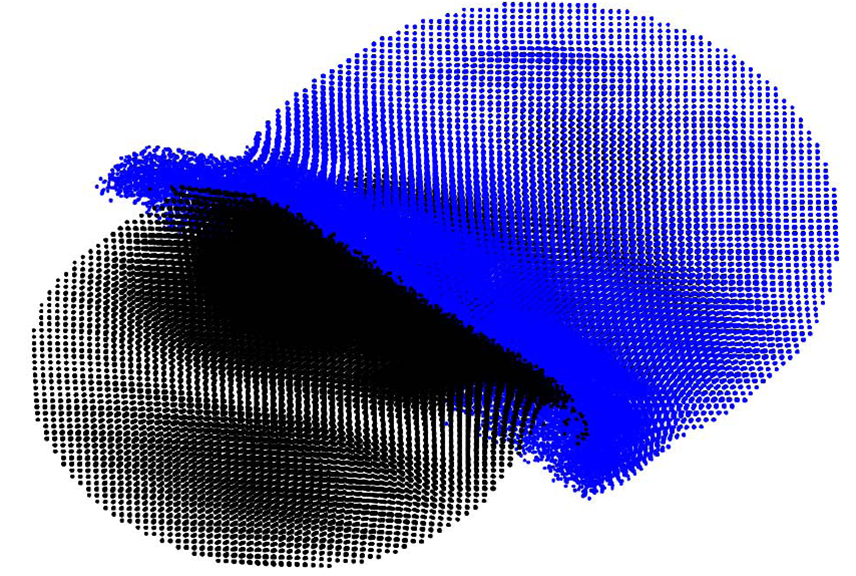}
\caption{Image of two colliding bodies of the size of Ceres computed with our SPH-code. 
The target has no water at all, the projectile has a layer of water on the surface (30 percent of its total mass).}
\label{ceres-water}
\end{figure}

As mentioned before the inner part of the primordial disk of planetesimals is dry so that
we need to find an explanation for the significant amount of water on Earth
inside the snow line. This border is located at
approximately 2 AU and it is by definition 
the minimum distance to the Sun for which water can be present in the liquid
state and also as ice. 
The collision 
behavior (see Fig.\ref{ceres-water} for an illustrative example) of small bodies coming from outside the snow line regarding their
water content is crucial to answer this question. They have to be 
modeled with specially designed programs like our SPH (Smoothed Particle Hydrodynamics) code, which will be discussed in detail in
the next chapter. In these simulations we fully take into account
the complicated process of (giant) collisions. For our N-body integrations on the other hand we just
assume perfect inelastic merging of two colliding bodies and determine the respective water loss at this
stage. The outcome of one such scenario as an example is shown in Fig.\ref{g21}.
We can see that planet {\bf A} with a mass of $\sim 1/3 M_{Earth}$ has a water content of at least two oceans, which
is in accordance with the estimated 4-8 oceans (surface and mantle combined) 
of present-day Earth. Rather surprising is the water content of the
innermost planet {\bf B}, located at Mercury's position but, 
we did not take into account the close proximity
to the central star, which may have made Mercury dry. Many questions are still open
and we are still far away from understanding the full 
complexity of planet formation, especially in connection to planet Earth, the 'BLUE
DOT' in our Solar System.
    
\section{Simulating collisions with SPH}

It is well established that planet formation results from a sequence of collision events between protoplanetary bodies of different sizes over an extended period of time. This collisional growth of planetesimals is investigated by means of dynamical evolution studies that are based on N-body simulations. The underlying collision model that leads to growth is typically simplified and based on conserving linear momentum in perfect inelastic merging \citep[e.g., ][]{dvoegg12, izides13, lunobr11} or deploying simple fragmentation models \citep{aleagn98}. As this approach neglects potential fragmentation in partial accretion and hit-and-run encounters, it misses the actual physics of collisions \citep{agncan99,asp10}. Rather, the resulting configuration after a collision depends on the involved bodies' mass ratios, encounter velocities, and impact angles. It can be categorized in one of the four regimes efficient accretion/perfect merging, partial accretion, hit-and-run, and erosion and disruption \citep{asp10}. \citet{agnasp04} for example, study giant collisions' accretion efficiency by simulating collisions using SPH (projectile and target masses of 0.1\,$M_\oplus\/$); \citet{marste09} extend disruption criteria for giant impacts up to masses of 10\,$M_\oplus\/$. As a consequence, one might expect consistently overestimated masses of the surviving bodies as supported e.g., by \citet{aleagn98} who investigated different simplified merging models and find the same number of big bodies on similar orbits, but depending on the merging assumption the bodies' masses increase towards higher merging rates. According to the computing power available at that time, \citet{aleagn98} simulate in 2D for up to $10^4\,\mathrm{yr}\/$ and deploy a very rudimentary fragmentation model which, however, predicts up to four surviving fragments and a remaining core.
 \citet{kokgen10} aim for a merging criterion and derive an empirical formula for the critical impact velocity that discriminates merging and hit-and-run collisions. They base their collision data on SPH collision simulations (resolution 20k--60k SPH particles). In simulations of protoplanetary systems for $3\times\,10^8\,\mathrm{yr}\/$ they find that only about half of the collisions lead to accretion. Nevertheless, no significant influence of the merging model on planetary growth timescales and the masses of the largest planets is found.

\citet{agncan99} analyzed the bodies' angular momenta  and show that assuming perfectly inelastic merging cannot be sustained because it would lead to rotationally unstable bodies. Spin rates resulting from the criterion by \citet{kokgen10} are about 30\,\% lower than in the perfect merging case.

All in all, 40--50\% of giant collisions are found to result in a non-merged configuration \citep{agnasp04,genkok12,kokgen10}.

As we are interested not only in the formation of planets, but also in water transport we decided to perform collision simulations including water content on the colliding bodies and consider effects of different material strengths of rocky and icy components.

\subsection{Physical model}
\label{sect:physmod}

Most giant impact studies \citep[e.g.,][]{canbar13} use strengthless material based on self-gravity dominating the material's tensile strength beyond a certain object size \citep[400\,m in radius,][]{melrya97}. In \citet{maidvo13b} we investigate collisions of objects close to this size limit and compare strengthless material simulation (``hydro model'') to modeling material strength with associated elasto-plastic effects and damage/brittle failure (``solid model'') which we will outline in this chapter.

We use the \citet{til62} equation of state (EOS) as formulated in \citet{mel89} for modeling the bodies' materials. It depends on 10 material constants $\rho_0$, $A\/$, $B\/$, $a\/$, $b\/$, $\alpha\/$, $\beta\/$, $E_0$, $E_\mathrm{iv}\/$, and $E_\mathrm{cv}\/$ and distinguishes two domains depending upon the material specific energy $E\/$. In compressed regions with density $\rho \geq \rho_0$) and $E\/$ lower than the energy of incipient vaporization $E_\mathrm{iv}$ the EOS reads
\begin{align}\label{eq:Pl}
P = \left[ a + \frac{b}{1+E/(E_0 \eta^2)} \right]\rho E + A\mu_\mathrm{T} + B\mu_\mathrm{T}^2
\end{align}
where $\eta = \rho / \rho_0$ and $\mu_\mathrm{T} = \eta-1$. For expanded states ($E$ greater than the energy of complete vaporization $E_\mathrm{cv}$) the EOS is given by
\begin{align}\label{eq:Ph}
P = a\rho E + & \bigg[ \frac{b\rho E}{1+E/(E_0 \eta^2)} 
 + \frac{A\mu_\mathrm{T}}{e^{\beta(\rho_0/\rho-1)}} \bigg] e^{-\alpha(\rho_0/\rho-1)^2}.
\end{align}
Linear interpolation between the pressures obtained via~(\ref{eq:Pl}) and~(\ref{eq:Ph}) gives the $P\/$ in the partial vaporization regime $E_\mathrm{iv} < E < E_\mathrm{cv}$ (refer to \citet{mel89} for a more detailed description).

Conservation of mass, momentum and energy describe the dynamics
of a solid body according to the theory of continuum mechanics, cf.\ \citet{schspe07}.
The continuity equation provides conservation of mass and reads in Lagrangian form (Einstein notation)
\begin{align*}
\frac{\mathrm{d}\rho}{\mathrm{d}t} + \rho \frac{\partial
v^\alpha}{\partial x^\alpha} = 0.
\end{align*}
Momentum is conserved by
\begin{align*} \frac{\mathrm{d}
v^\alpha}{\mathrm{d} t} = \frac{1}{\rho} \frac{\partial \sigma^{\alpha
\beta}}{\partial x^\beta},\quad \sigma^{\alpha \beta} = -P \delta^{\alpha \beta} +
S^{\alpha \beta}
\end{align*}
with the stress tensor $\sigma^{\alpha \beta}$ that can be split between the pressure $P\/$ along its diagonal and and the deviatoric stress tensor $S^{\alpha \beta}$ ($\delta^{\alpha \beta}$ denotes the
Kronecker delta).
Energy conservation reads
\begin{align*}
\frac{\mathrm{d}E}{\mathrm{d}t} = -\frac{P}{\rho}\frac{\partial v^{\alpha}}{\partial x^\alpha} + \frac{1}{\rho}S^{\alpha \beta}\dot{\epsilon}^{\alpha \beta},
\end{align*}
where $\dot{\epsilon}^{\alpha \beta}$ denotes the strain rate tensor as given in~(\ref{eq:repsdot}).

While the EOS provides an analytic function for the pressure, the time evolution of the deviatoric stress tensor $S^{\alpha \beta}$ is needed to completely describe the dynamics of a solid body. We use Hooke's law to define the constitutive equation as
\begin{align*}
\frac{\mathrm{d}S^{\alpha \beta}}{\mathrm{d}t} = 2 \mu \left(
\dot{\epsilon}^{\alpha \beta} - \frac{1}{3} \delta^{\alpha \beta}
\dot{\epsilon}^{\gamma\gamma} \right) + S^{\alpha \gamma} R^{\gamma \beta} - R^{\alpha \gamma}S^{\gamma \beta}, 
\end{align*}
where $\mu\/$ represents the shear modulus and $\dot{\epsilon}^{\alpha \beta}\/$ the strain rate tensor. The last two terms depend on the rotation rate tensor $R^{\alpha \beta}\/$ and guarantee that the constitutive equations are independent from the material frame of reference.
$R^{\alpha \beta}\/$ and $\dot{\epsilon}^{\alpha \beta}\/$ are given by
\begin{align}
R^{\alpha \beta} = \frac{1}{2} \left( \frac{\partial v^\alpha}{\partial
x^\beta} - \frac{\partial v^\beta}{\partial x^\alpha} \right), \;
\dot{\epsilon}^{\alpha \beta} = \frac{1}{2} \left( \frac{\partial v^\alpha}{\partial
x^\beta} + \frac{\partial v^\beta}{\partial x^\alpha}\right).\label{eq:repsdot}
\end{align}
Equations (\ref{eq:Pl}) to (\ref{eq:repsdot}) describe the dynamics of an elastic solid body. We follow the approach by \citet{benasp94} to model plastic behavior and use the von Mises yield criterion where the material yield stress $Y$ limits the deviatoric stress. Hence, a transformed deviatoric stress $S^{\alpha \beta}_\mathrm{vM}$ is introduced:
\begin{align*}
S^{\alpha \beta}_\mathrm{vM} = \min \left[\frac{2\,Y^2}{3\,S^{\alpha \beta}S^{\alpha \beta}},1\right]\cdot S^{\alpha \beta}.
\end{align*}

We model tensile failure (fracture) by implementing a damage model based on the fragmentation model of \citet{grakip80} and the SPH implementation by \citet{benasp94}. The model is based on assigning flaws to SPH particles that are activated once the strain exceeds a certain level $\epsilon\/$. Activated flaws convert into cracks. The ratio of cracks to flaws defines the damage value $d\in [0,1]\/$ assigned to each SPH particle. The deviatoric stress and tension decrease proportionally to $1-d\/$ and vanish for completely damaged material ($d=1\/$). Initially, the number $n\/$ of flaws that can be activated by a strain level of $\epsilon$ are distributed according to a Weibull distribution with material parameters $k$ and $m$: $n(\epsilon)=k\,\epsilon^m$ \citep{wei39}.

\begin{table*}
\caption{Tillotson EOS parameters, vaporization energy levels, shear modulus $\mu\/$, and yield stress $Y\/$, cf.\ \citet{benasp99}. $A=B\/$ is set equal to the bulk modulus.\label{t:EOSpar}}
\begin{ruledtabular}
\begin{tabular}{l*{10}{r}rl}
 \tstrut & \tc{$\rho_0$} & \tc{$A$} & \tc{$B$} & \tc{$E_0$} & \tc{$E_\mathrm{iv}$}
         & \tc{$E_\mathrm{cv}$} & \tc{\multirow{2}{*}{$a$}}
                                & \tc{\multirow{2}{*}{$b$}}
                                & \tc{\multirow{2}{*}{$\alpha$}} 
                                & \tc{\multirow{2}{*}{$\beta$}}
                                & \tc{$\mu$} & \tc{$Y$} \\
       & (kg/m$^3$) & (GPa) & (GPa) & (MJ/kg) & (MJ/kg) & (MJ/kg) & & & & 
       & (GPa)      & (GPa) \\
\hline
Basalt & 2700 & 26.7\hspace{\digit}  & 26.7\hspace{\digit}  & 487 & 4.72\hspace{\digit}  & 18.2\hspace{\digit}  & 0.5 & 1.50 &  5.0 & \tstrut 5.0
& 22.7       & \; 3.5 \\
Ice    & \hspace{\digit}917 & 9.47 & 9.47 & \hspace{\digit}10 & 0.773 & 3.04 & 0.3 & 0.1  & 10.0 & 5.0 & 2.8        & \; 1 \\
\end{tabular}
\end{ruledtabular}
\end{table*}
Table~\ref{t:EOSpar} gives the Tillotson EOS parameters, the shear modulus $\mu\/$ and yield stress $Y\/$ which we use in our simulations. The values are adopted from \citet{benasp99} following their reasoning on approximating $\rho_0$, setting $A\/$ equal to the bulk modulus, and $B=A\/$. 

We use Weibull flaw distribution parameters of basalt based on directly measured data by \citet{nakmic07} and for ice we adopt the values mentioned in \citet{lanahr84}: $m_\mathrm{basalt}=16$, $k_\mathrm{basalt}=10^{61}\,\mathrm{m}^{-3}$, $m_\mathrm{ice}=9.1$, $k_\mathrm{ice}=10^{46}\,\mathrm{m}^{-3}$.

\subsection{Fragmentation and water loss}

The presented study was performed with our own smoothed particle hydrodynamics (SPH) code which is based on the physics described in the previous section with the addition of self-gravity and a tensorial correction ensuring first-order consistency as described in \citet{schspe07}. The SPH code has been introduced in \citet{sch05,maisch13,maidvo13b} and used for simulating high-resolution surface impacts \citep{maidvo14,maihag14} and planetesimal-to-protoplanet collisions \citep{maidvo13,maidvo13b,maidvo14b}.

Our scenarios presented here represent two colliding objects that have a mass of $M_\mathrm{Ceres}\/$ each. The target is assumed to consist of a rocky basalt core (70\,mass-\%) and a mantle/shell of water ice (30\,mass-\%) whereas the projectile is purely rocky (basalt). In total we simulated 42 scenarios parameterized by collision velocities $v_0\/$ between 0.95 and 5.88 two-body escape velocities $v_\mathrm{esc}\/$ and impact angles $\alpha\/$ between $0^\circ\/$ (head-on) and a flyby. As we are interested in an overall fragmentation and water transport pattern we use a relatively low resolution of 20k SPH particles.

\subsubsection{Collision outcome}

Due to the typically high degree of fragmentation after impacts in the solid model and in order to get statistically relevant physical fragment  properties we limit our study to fragments with masses $m_\mathrm{frag}\/$ corresponding to at least 20 basalt SPH particles. Given the scenarios' single basalt SPH particle mass of $1.2\times 10^{17}\,\mathrm{kg}\/$ this results in the condition $m_\mathrm{frag}\ge 2.4\times 10^{18}\,\mathrm{kg}\/$ while the remaining mass will be considered as lost debris. This limits the resolution of our results. Table~\ref{tab:nfrags} states the parameters of the 42 different collision configurations (impact velocity and angle) simulated in \citet{maidvo13b} along with the resulting numbers \NF of such ``significant fragments'' after the collision. Each $\alpha\/$-\NF column pair corresponds to one initial impact parameter. The bodies are started five mean diameters apart to allow the particle distribution to settle due to mutual tidal forces. Because of mutual gravitational interaction different encounter speeds result in different orbital deflection and hence different actual impact angles $\alpha\/$. In case of the largest initial impact parameter (rightmost column in Tab.~\ref{tab:nfrags}) only the slowest encounter velocity leads to an impact; otherwise the scenario results in a flyby and hence cannot be assigned an impact angle. The latter show two surviving fragments which are exactly the original bodies.
\begin{table*}
\caption{Number of significant surviving fragments \NF after 2000\,min simulation time. The scenario parameters are the collision velocity $v_0\/$ given in units of $v_\mathrm{esc}$ (two-body escape velocity) and the impact angle $\alpha\/$. The $v_0\/$ (averaged) and $\alpha\/$ values are taken from \citet{maidvo13b}; a missing impact angle indicates a flyby scenario (see text).\label{tab:nfrags}}
\begin{ruledtabular}
\begin{tabular}{r@{\qquad}*{5}{rc@{\qquad}}rc}
\tstrut$v_0$ &	\tc{$\alpha$} & \NF & \tc{$\alpha$} & \NF & \tc{$\alpha$} & \NF & \tc{$\alpha$} & \NF & \tc{$\alpha$} & \NF & \tc{$\alpha$} & \NF \\
\tstrut$[v_\mathrm{esc}]$ & \tc{[$^\circ$]} & & \tc{[$^\circ$]} & & \tc{[$^\circ$]} & & \tc{[$^\circ$]} & & \tc{[$^\circ$]} & & \tc{[$^\circ$]} \\
\hline						
\tstrut
0.95 & 0 &\dd 1 & 14 &\dd 1 & 23 &\dd 1 & 25 &\dd 1 & 31 &\dd 1 & 62     &\dd 7 \\
1.32 & 0 &\dd 1 & 11 &\dd 1 & 21 &\dd 1 & 40 &\dd 1 & 48 &\dd 2 & \tc{-} &\dd 2 \\
1.36 & 0 &\dd 1 & 11 &\dd 1 & 20 &\dd 1 & 40 &\dd 2 & 48 &\dd 2 & \tc{-} &\dd 2  \\
2.12 & 0 &\dd 1 & 12 &\dd 1 & 25 &\dd 5 & 50 &\dd 2 & 62 &\dd 6 & \tc{-} &\dd 2  \\
3.04 & 0 &   39 & 12 &   52 & 25 &   22 & 53 &\dd 2 & 67 &\dd 2 & \tc{-} &\dd 2  \\
3.97 & 0 &   41 & 13 &   67 & 27 &   35 & 55 &\dd 2 & 72 &\dd 2 & \tc{-} &\dd 2  \\
5.88 & 0 &   45 & 13 &   60 & 28 &   61 & 58 &\dd 2 & 72 &\dd 2 & \tc{-} &\dd 2  \\
\end{tabular}
\end{ruledtabular}
\end{table*}

\begin{figure*}
 \includegraphics[width=0.62\textwidth]{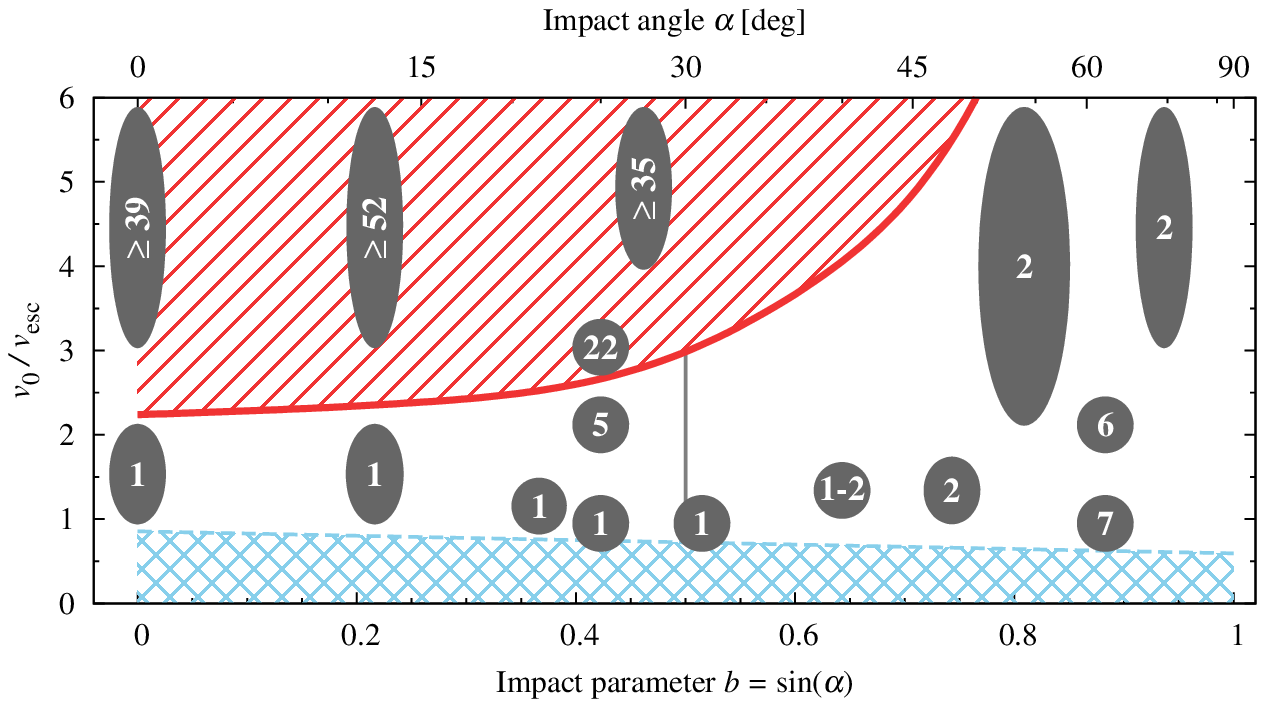} 
 \caption{Number of significant surviving fragments for the scenarios' collision velocities $v_0\/$ given in units of $v_\mathrm{esc}$ (two-body escape velocity) and impact angles $\alpha\/$. Different areas in the graph correspond to erosion (red, shaded), partial accretion (white, $\alpha \lesssim 30^\circ\/$), hit-and-run (white, $\alpha \gtrsim 30^\circ\/$), and perfect merging (blue, hashed) as given by the analytic model for strengthless planet collisions in \citet{leiste12}, Fig.~11A. See text for discussion.\label{fig:collmap}}
\end{figure*}

Fig.~\ref{fig:collmap} relates our fragmentation results to an analytic collision outcome model for strengthless planets presented by \citet{leiste12}. Overall, the \NF numbers correspond to the model, with some exceptions at the boundaries of the collision outcome regimes. They suggest that the onset of the hit-and-run regime actually happens at a higher impact angle than predicted by the strengthless model. At higher angles the \NF$=6$ bubble actually represents a hit-and-run with four very small fragments which is in accordance to the analytical model, whereas the \NF$=7$ outcome is actually a merging event with 6 very small fragments and suggests the merging limit to be slightly too low in the analytic model. Please refer to \citet{maidvo13b,maidvo14b} for a more thorough discussion of the detailed outcomes.

\subsubsection{Water content}

\begin{figure*}
 \includegraphics[width=0.62\textwidth]{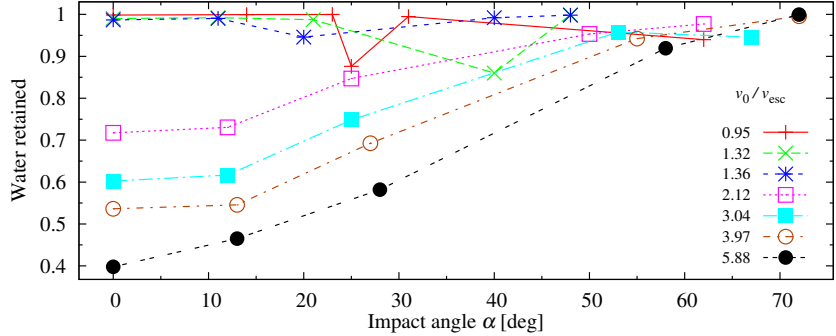} 
 \caption{Cumulated water retained in significant fragments after the collision. A value of one means all water in the system is on the surviving significant fragments. The different curves correspond to different collision velocities $v_0\/$ in units of the two-body escape velocity $v_\mathrm{esc}\/$. See text for discussion.\label{fig:water}}
\end{figure*}

Tracking overall water content retained in significant fragments after the collision and comparing to the water content on the original bodies gives an estimate on the loss of volatiles such as water ice. Figure~\ref{fig:water} shows how much of the water is still on significant fragments after the collision. The remaining water is on the debris and lost in our model. For $\alpha \lesssim 20^\circ\/$ and $v_0 \lesssim 1.3\,v_\mathrm{esc}\/$ most of the water ice stays on the survivor, in strongly inclined hit-and-run scenarios more than 80\,\% of the water stays as well. In general more water is lost for ``more violent'' impacts---an increasing amount of water ice is lost in debris for small collision angles and high velocities with a more significant dependency on velocity than on the angle.

Notable features in Fig.~\ref{fig:water} include downward spikes occurring in the low velocity scenarios ($v_0 \le 1.36\,v_\mathrm{esc}\/$). A more thorough investigation reveals that these are caused by a single surviving body spinning rapidly losing debris including large portions of its surface water ice. A more detailed discussion of fragment counts and individual scenario descriptions is available in \citet{maidvo13b}.

\subsubsection{Future work}

Based on earlier results of simulating small to mid scale collisions of planetesimals at moderate energies \citep{maidvo13b,maidvo14b} we presented first conclusions on fragmentation and volatile loss. Rather than analyzing total fragment counts we identified surviving ``significant fragments'' and their volatile content. While the collision outcomes qualitatively agree with analytic giant collision outcome models of strengthless planets \citep[as given e.g., in ][]{leiste12}, we observe some minor shifts of boundaries between the merging, erosion, and hit-and-run regimes. We further observe significant water loss---up to 60\,\%---for faster and/or less inclined impacts. This might have a major impact on future planet formation simulations as most of them currently rely on a 100\,\% water transfer whenever a collision occurs. Incorporating a realistic model of volatile transport and loss may result in reduced water abundances of formed terrestrial planets by a factor of 5--10 \citep{hagray07}. Envisioned future studies will include fragment dynamics determining their ability to escape the system's Hill sphere connected with working towards a comprehensive model for the fate of volatiles such as water ice and including it in N-body simulations for planet formation.

\appendix*
\section{Appendix: The Discovery of Exoplanets}

Up to now we have knowledge of 1164 planetary systems, where 473 of them host multiple companions, giving a total of 1855 known extrasolar planets (\emph{www.exoplanet.eu}, last updated 2014).
The believe in other worlds similar to our own dates back to the ancient
Greeks like  Democritus ($\sim$ 460-370 B.C.) and Epicurus (341-270 B.C.).
2000 years later, in the sixteenth century, the Italian philosopher
Giordano Bruno,  who was a supporter of the Copernican theory, put forward 
the idea that the stars are similar to the Sun and could also be accompanied by
planets.

It is interesting to note that the first confirmed planets, orbiting a pulsar \citep{1992Natur.355..145W}, were rather small
compared to most later discoveries. Only 5 years later the
first planet around a main sequence star, a hot Jupiter with a 4.2-day orbit (51 Pegasi), was discovered by \citet{1995Natur.378..355M}.

In the following we summarize the different techniques for discovering extrasolar planets:

\begin{itemize}
\item  radial velocity technique (RV)
\item transit photometry (TP)
\item microlensing (ML)
\item direct imaging (DI)
\end{itemize}

The first three are indirect methods, making it impossible to
directly separate the orbiting planet from the host star. Furthermore one 
has to distinguish
ground based detections and observations via space telescopes, a method up to now only applied to the TP technique.

\begin{enumerate}

\item {\bf RV}   
This method, also called Doppler technique, uses the Doppler effect for
analyzing the motion and properties of the star and planet. Because both, 
the planet and the star, are orbiting their common center of mass, spectral lines are shifted versus shorter 
wavelengths (blue) when the host
star moves towards the observer. When the planet moves towards the observer on the contrary, the star is moving
away, and consequently the spectral lines of the star are shifted versus longer
wavelengths (red). This change happens periodically since the planet orbits the star with a certain period. This allows us to obtain a so-called RV curve, where one can very well determine the period and even the orbital eccentricity from.

\item {\bf TP}
Like Sun, Moon and Earth are aligned during a solar eclipse, observer, planet and its host star are in one straight line during a TP event. The extrasolar planet is moving in front of the stellar disk just like
we can observe it on Earth for Mercury and Venus every now and then. 
Knowing the real diameter of the star via spectroscopic observations and also in addition its RV curve (typically from follow-up observations) one can determine the period of the orbit, the planet's diameter, the orbital eccentricity and in some instances also the inclination. 
Having obtained these values it is straight forward to estimate the exoplanet's mass and bulk density with the aid of the $3^{rd}$ law of Kepler.
These parameters are important because at least for most examples they tell us whether it is rather a gaseous planet or a rocky one.

\item {\bf ML}
 When a distant star is almost aligned with a massive compact
foreground star hosting a planet, the distortion of light due to its 
gravitational field may lead to two unresolved images and thus to an
observable magnification. This transient brightening depends on the mass 
of the object which acts as a lens. It is important to mention that this is always a unique event because the almost alignments of the observer on Earth and the
two other objects are quite rare events. In addition the time of possible
observations depends on the relative proper motion between the background
'source' and the foreground 'lens' object. 

\item {\bf DI} Because of the big difference in luminosity between the star
and a planet this technique has so far been successful only for big planets relatively far
away from their host star.

\end{enumerate}

For the last ten years photometric observations from space are carried out as well: the first larger instrument was the CoRoT (COnvection, ROtation 
and planetary Transits) satellite, started in 2006 by ESA and was working well
for about 8 years. With a 27cm
telescope on board thousands of stars were investigated, leading to hundreds of 
planetary candidates and about 40 confirmed exoplanets.
NASA's KEPLER satellite observes with a 95cm telescope and was launched in
2009.
It is still working and the bigger diameter of the telescope allowed for
observations of hundreds of thousands of stars, resulting in more than 1000 planetary candidates and several hundreds of confirmed planets by later measurements from the ground. When speaking about candidates and confirmed planets it should be mentioned
that observations from space alone are not enough, ground based observations, mainly spectroscopic ones for obtaining additional RV curves, need to be
done. With the TP method one may be observing the dimming of light 
caused by star spots, an apparently close by but indeed far away background star or even a close binary, leading to a false-positive planet detection.

To summarize one can say that the last twenty years of research on extrasolar
planets opened a new and exciting chapter in the field of astronomy, and also for
the human civilization as a whole, trying to answer the big question: \emph{ARE WE ALONE?}



\section{Acknowledgements}
This research is produced as part of the FWF Austrian Science Fund project S~11603-N16. We also acknowledge support from FWF project P23810-N16.
In part the calculations for this work were performed on the hpc-bw-cluster---we gratefully thank the bwGRiD project\footnote{bwGRiD (http://www.bw-grid.de), member of the German D-Grid initiative, funded by the Ministry for Education and Research (Bundesministerium fuer Bildung und Forschung) and the Ministry for Science, Research and Arts Baden-Wuerttemberg (Ministerium fuer Wissenschaft, Forschung und Kunst Baden-Wuerttemberg).} for the computational resources.
 

\bibliography{./references_thomas.bib,./references_rudi.bib}

\begin{thebibliography}{50}%
\makeatletter
\providecommand \@ifxundefined [1]{%
 \@ifx{#1\undefined}
}%
\providecommand \@ifnum [1]{%
 \ifnum #1\expandafter \@firstoftwo
 \else \expandafter \@secondoftwo
 \fi
}%
\providecommand \@ifx [1]{%
 \ifx #1\expandafter \@firstoftwo
 \else \expandafter \@secondoftwo
 \fi
}%
\providecommand \natexlab [1]{#1}%
\providecommand \enquote  [1]{``#1''}%
\providecommand \bibnamefont  [1]{#1}%
\providecommand \bibfnamefont [1]{#1}%
\providecommand \citenamefont [1]{#1}%
\providecommand \href@noop [0]{\@secondoftwo}%
\providecommand \href [0]{\begingroup \@sanitize@url \@href}%
\providecommand \@href[1]{\@@startlink{#1}\@@href}%
\providecommand \@@href[1]{\endgroup#1\@@endlink}%
\providecommand \@sanitize@url [0]{\catcode `\\12\catcode `\$12\catcode
  `\&12\catcode `\#12\catcode `\^12\catcode `\_12\catcode `\%12\relax}%
\providecommand \@@startlink[1]{}%
\providecommand \@@endlink[0]{}%
\providecommand \url  [0]{\begingroup\@sanitize@url \@url }%
\providecommand \@url [1]{\endgroup\@href {#1}{\urlprefix }}%
\providecommand \urlprefix  [0]{URL }%
\providecommand \Eprint [0]{\href }%
\providecommand \doibase [0]{http://dx.doi.org/}%
\providecommand \selectlanguage [0]{\@gobble}%
\providecommand \bibinfo  [0]{\@secondoftwo}%
\providecommand \bibfield  [0]{\@secondoftwo}%
\providecommand \translation [1]{[#1]}%
\providecommand \BibitemOpen [0]{}%
\providecommand \bibitemStop [0]{}%
\providecommand \bibitemNoStop [0]{.\EOS\space}%
\providecommand \EOS [0]{\spacefactor3000\relax}%
\providecommand \BibitemShut  [1]{\csname bibitem#1\endcsname}%
\let\auto@bib@innerbib\@empty
\bibitem [{\citenamefont {{Armitage}}(2007)}]{2007astro.ph..1485A}%
  \BibitemOpen
  \bibfield  {author} {\bibinfo {author} {\bibfnamefont {P.~J.}\ \bibnamefont
  {{Armitage}}},\ }\href@noop {} {\bibfield  {journal} {\bibinfo  {journal}
  {ArXiv Astrophysics e-prints}\ } (\bibinfo {year} {2007})},\ \Eprint
  {http://arxiv.org/abs/astro-ph/0701485} {astro-ph/0701485} \BibitemShut
  {NoStop}%
\bibitem [{\citenamefont {{Walsh}}\ \emph {et~al.}(2011)\citenamefont
  {{Walsh}}, \citenamefont {{Morbidelli}}, \citenamefont {{Raymond}},
  \citenamefont {{O'Brien}},\ and\ \citenamefont
  {{Mandell}}}]{2011Natur.475..206W}%
  \BibitemOpen
  \bibfield  {author} {\bibinfo {author} {\bibfnamefont {K.~J.}\ \bibnamefont
  {{Walsh}}}, \bibinfo {author} {\bibfnamefont {A.}~\bibnamefont
  {{Morbidelli}}}, \bibinfo {author} {\bibfnamefont {S.~N.}\ \bibnamefont
  {{Raymond}}}, \bibinfo {author} {\bibfnamefont {D.~P.}\ \bibnamefont
  {{O'Brien}}}, \ and\ \bibinfo {author} {\bibfnamefont {A.~M.}\ \bibnamefont
  {{Mandell}}},\ }\href {\doibase 10.1038/nature10201} {\bibfield  {journal}
  {\bibinfo  {journal} {\nat}\ }\textbf {\bibinfo {volume} {475}},\ \bibinfo
  {pages} {206} (\bibinfo {year} {2011})},\ \Eprint
  {http://arxiv.org/abs/1201.5177} {arXiv:1201.5177 [astro-ph.EP]} \BibitemShut
  {NoStop}%
\bibitem [{\citenamefont {{Hansen}}(2009)}]{2009ApJ...703.1131H}%
  \BibitemOpen
  \bibfield  {author} {\bibinfo {author} {\bibfnamefont {B.~M.~S.}\
  \bibnamefont {{Hansen}}},\ }\href {\doibase 10.1088/0004-637X/703/1/1131}
  {\bibfield  {journal} {\bibinfo  {journal} {\apj}\ }\textbf {\bibinfo
  {volume} {703}},\ \bibinfo {pages} {1131} (\bibinfo {year} {2009})},\ \Eprint
  {http://arxiv.org/abs/0908.0743} {arXiv:0908.0743 [astro-ph.EP]} \BibitemShut
  {NoStop}%
\bibitem [{\citenamefont {{Izidoro}}\ \emph {et~al.}(2014)\citenamefont
  {{Izidoro}}, \citenamefont {{Haghighipour}}, \citenamefont {{Winter}},\ and\
  \citenamefont {{Tsuchida}}}]{2014ApJ...782...31I}%
  \BibitemOpen
  \bibfield  {author} {\bibinfo {author} {\bibfnamefont {A.}~\bibnamefont
  {{Izidoro}}}, \bibinfo {author} {\bibfnamefont {N.}~\bibnamefont
  {{Haghighipour}}}, \bibinfo {author} {\bibfnamefont {O.~C.}\ \bibnamefont
  {{Winter}}}, \ and\ \bibinfo {author} {\bibfnamefont {M.}~\bibnamefont
  {{Tsuchida}}},\ }\href {\doibase 10.1088/0004-637X/782/1/31} {\bibfield
  {journal} {\bibinfo  {journal} {\apj}\ }\textbf {\bibinfo {volume} {782}},\
  \bibinfo {eid} {31} (\bibinfo {year} {2014})},\ \Eprint
  {http://arxiv.org/abs/1312.3959} {arXiv:1312.3959 [astro-ph.EP]} \BibitemShut
  {NoStop}%
\bibitem [{\citenamefont {{Jin}}\ \emph {et~al.}(2008)\citenamefont {{Jin}},
  \citenamefont {{Arnett}}, \citenamefont {{Sui}},\ and\ \citenamefont
  {{Wang}}}]{2008ApJ...674L.105J}%
  \BibitemOpen
  \bibfield  {author} {\bibinfo {author} {\bibfnamefont {L.}~\bibnamefont
  {{Jin}}}, \bibinfo {author} {\bibfnamefont {W.~D.}\ \bibnamefont {{Arnett}}},
  \bibinfo {author} {\bibfnamefont {N.}~\bibnamefont {{Sui}}}, \ and\ \bibinfo
  {author} {\bibfnamefont {X.}~\bibnamefont {{Wang}}},\ }\href {\doibase
  10.1086/529375} {\bibfield  {journal} {\bibinfo  {journal} {\apjl}\ }\textbf
  {\bibinfo {volume} {674}},\ \bibinfo {pages} {L105} (\bibinfo {year}
  {2008})}\BibitemShut {NoStop}%
\bibitem [{\citenamefont {{Raymond}}\ \emph {et~al.}(2009)\citenamefont
  {{Raymond}}, \citenamefont {{O'Brien}}, \citenamefont {{Morbidelli}},\ and\
  \citenamefont {{Kaib}}}]{2009Icar..203..644R}%
  \BibitemOpen
  \bibfield  {author} {\bibinfo {author} {\bibfnamefont {S.~N.}\ \bibnamefont
  {{Raymond}}}, \bibinfo {author} {\bibfnamefont {D.~P.}\ \bibnamefont
  {{O'Brien}}}, \bibinfo {author} {\bibfnamefont {A.}~\bibnamefont
  {{Morbidelli}}}, \ and\ \bibinfo {author} {\bibfnamefont {N.~A.}\
  \bibnamefont {{Kaib}}},\ }\href {\doibase 10.1016/j.icarus.2009.05.016}
  {\bibfield  {journal} {\bibinfo  {journal} {\icarus}\ }\textbf {\bibinfo
  {volume} {203}},\ \bibinfo {pages} {644} (\bibinfo {year} {2009})},\ \Eprint
  {http://arxiv.org/abs/0905.3750} {arXiv:0905.3750 [astro-ph.EP]} \BibitemShut
  {NoStop}%
\bibitem [{\citenamefont {{Raymond}}\ \emph {et~al.}(2014)\citenamefont
  {{Raymond}}, \citenamefont {{Kokubo}}, \citenamefont {{Morbidelli}},
  \citenamefont {{Morishima}},\ and\ \citenamefont
  {{Walsh}}}]{2014prpl.conf..595R}%
  \BibitemOpen
  \bibfield  {author} {\bibinfo {author} {\bibfnamefont {S.~N.}\ \bibnamefont
  {{Raymond}}}, \bibinfo {author} {\bibfnamefont {E.}~\bibnamefont {{Kokubo}}},
  \bibinfo {author} {\bibfnamefont {A.}~\bibnamefont {{Morbidelli}}}, \bibinfo
  {author} {\bibfnamefont {R.}~\bibnamefont {{Morishima}}}, \ and\ \bibinfo
  {author} {\bibfnamefont {K.~J.}\ \bibnamefont {{Walsh}}},\ }\href {\doibase
  10.2458/azu_uapress_9780816531240-ch026} {\bibfield  {journal} {\bibinfo
  {journal} {Protostars and Planets VI}\ ,\ \bibinfo {pages} {595}} (\bibinfo
  {year} {2014})},\ \Eprint {http://arxiv.org/abs/1312.1689} {arXiv:1312.1689
  [astro-ph.EP]} \BibitemShut {NoStop}%
\bibitem [{\citenamefont {{Agnor}}\ and\ \citenamefont
  {{Asphaug}}(2004{\natexlab{a}})}]{2004ApJ...613L.157A}%
  \BibitemOpen
  \bibfield  {author} {\bibinfo {author} {\bibfnamefont {C.}~\bibnamefont
  {{Agnor}}}\ and\ \bibinfo {author} {\bibfnamefont {E.}~\bibnamefont
  {{Asphaug}}},\ }\href {\doibase 10.1086/425158} {\bibfield  {journal}
  {\bibinfo  {journal} {\apjl}\ }\textbf {\bibinfo {volume} {613}},\ \bibinfo
  {pages} {L157} (\bibinfo {year} {2004}{\natexlab{a}})}\BibitemShut {NoStop}%
\bibitem [{\citenamefont {{Marcus}}\ \emph {et~al.}(2010)\citenamefont
  {{Marcus}}, \citenamefont {{Sasselov}}, \citenamefont {{Stewart}},\ and\
  \citenamefont {{Hernquist}}}]{2010ApJ...719L..45M}%
  \BibitemOpen
  \bibfield  {author} {\bibinfo {author} {\bibfnamefont {R.~A.}\ \bibnamefont
  {{Marcus}}}, \bibinfo {author} {\bibfnamefont {D.}~\bibnamefont
  {{Sasselov}}}, \bibinfo {author} {\bibfnamefont {S.~T.}\ \bibnamefont
  {{Stewart}}}, \ and\ \bibinfo {author} {\bibfnamefont {L.}~\bibnamefont
  {{Hernquist}}},\ }\href {\doibase 10.1088/2041-8205/719/1/L45} {\bibfield
  {journal} {\bibinfo  {journal} {\apjl}\ }\textbf {\bibinfo {volume} {719}},\
  \bibinfo {pages} {L45} (\bibinfo {year} {2010})},\ \Eprint
  {http://arxiv.org/abs/1007.3212} {arXiv:1007.3212 [astro-ph.EP]} \BibitemShut
  {NoStop}%
\bibitem [{\citenamefont {{Agnor}}\ and\ \citenamefont
  {{Lin}}(2012)}]{2012ApJ...745..143A}%
  \BibitemOpen
  \bibfield  {author} {\bibinfo {author} {\bibfnamefont {C.~B.}\ \bibnamefont
  {{Agnor}}}\ and\ \bibinfo {author} {\bibfnamefont {D.~N.~C.}\ \bibnamefont
  {{Lin}}},\ }\href {\doibase 10.1088/0004-637X/745/2/143} {\bibfield
  {journal} {\bibinfo  {journal} {\apj}\ }\textbf {\bibinfo {volume} {745}},\
  \bibinfo {eid} {143} (\bibinfo {year} {2012})},\ \Eprint
  {http://arxiv.org/abs/1110.5042} {arXiv:1110.5042 [astro-ph.EP]} \BibitemShut
  {NoStop}%
\bibitem [{\citenamefont {{Morbidelli}}\ \emph {et~al.}(2012)\citenamefont
  {{Morbidelli}}, \citenamefont {{Lunine}}, \citenamefont {{O'Brien}},
  \citenamefont {{Raymond}},\ and\ \citenamefont
  {{Walsh}}}]{2012AREPS..40..251M}%
  \BibitemOpen
  \bibfield  {author} {\bibinfo {author} {\bibfnamefont {A.}~\bibnamefont
  {{Morbidelli}}}, \bibinfo {author} {\bibfnamefont {J.~I.}\ \bibnamefont
  {{Lunine}}}, \bibinfo {author} {\bibfnamefont {D.~P.}\ \bibnamefont
  {{O'Brien}}}, \bibinfo {author} {\bibfnamefont {S.~N.}\ \bibnamefont
  {{Raymond}}}, \ and\ \bibinfo {author} {\bibfnamefont {K.~J.}\ \bibnamefont
  {{Walsh}}},\ }\href {\doibase 10.1146/annurev-earth-042711-105319} {\bibfield
   {journal} {\bibinfo  {journal} {Annual Review of Earth and Planetary
  Sciences}\ }\textbf {\bibinfo {volume} {40}},\ \bibinfo {pages} {251}
  (\bibinfo {year} {2012})},\ \Eprint {http://arxiv.org/abs/1208.4694}
  {arXiv:1208.4694 [astro-ph.EP]} \BibitemShut {NoStop}%
\bibitem [{\citenamefont {{Lineweaver}}\ and\ \citenamefont
  {{Chopra}}(2012)}]{2012AREPS..40..597L}%
  \BibitemOpen
  \bibfield  {author} {\bibinfo {author} {\bibfnamefont {C.~H.}\ \bibnamefont
  {{Lineweaver}}}\ and\ \bibinfo {author} {\bibfnamefont {A.}~\bibnamefont
  {{Chopra}}},\ }\href {\doibase 10.1146/annurev-earth-042711-105531}
  {\bibfield  {journal} {\bibinfo  {journal} {Annual Review of Earth and
  Planetary Sciences}\ }\textbf {\bibinfo {volume} {40}},\ \bibinfo {pages}
  {597} (\bibinfo {year} {2012})}\BibitemShut {NoStop}%
\bibitem [{\citenamefont {{G{\"u}del}}\ \emph {et~al.}(2014)\citenamefont
  {{G{\"u}del}}, \citenamefont {{Dvorak}}, \citenamefont {{Erkaev}},
  \citenamefont {{Kasting}}, \citenamefont {{Khodachenko}}, \citenamefont
  {{Lammer}}, \citenamefont {{Pilat-Lohinger}}, \citenamefont {{Rauer}},
  \citenamefont {{Ribas}},\ and\ \citenamefont {{Wood}}}]{2014prpl.conf..883G}%
  \BibitemOpen
  \bibfield  {author} {\bibinfo {author} {\bibfnamefont {M.}~\bibnamefont
  {{G{\"u}del}}}, \bibinfo {author} {\bibfnamefont {R.}~\bibnamefont
  {{Dvorak}}}, \bibinfo {author} {\bibfnamefont {N.}~\bibnamefont {{Erkaev}}},
  \bibinfo {author} {\bibfnamefont {J.}~\bibnamefont {{Kasting}}}, \bibinfo
  {author} {\bibfnamefont {M.}~\bibnamefont {{Khodachenko}}}, \bibinfo {author}
  {\bibfnamefont {H.}~\bibnamefont {{Lammer}}}, \bibinfo {author}
  {\bibfnamefont {E.}~\bibnamefont {{Pilat-Lohinger}}}, \bibinfo {author}
  {\bibfnamefont {H.}~\bibnamefont {{Rauer}}}, \bibinfo {author} {\bibfnamefont
  {I.}~\bibnamefont {{Ribas}}}, \ and\ \bibinfo {author} {\bibfnamefont
  {B.~E.}\ \bibnamefont {{Wood}}},\ }\href {\doibase
  10.2458/azu_uapress_9780816531240-ch038} {\bibfield  {journal} {\bibinfo
  {journal} {Protostars and Planets VI}\ ,\ \bibinfo {pages} {883}} (\bibinfo
  {year} {2014})},\ \Eprint {http://arxiv.org/abs/1407.8174} {arXiv:1407.8174
  [astro-ph.EP]} \BibitemShut {NoStop}%
\bibitem [{\citenamefont {{Chambers}}(1999)}]{1999MNRAS.304..793C}%
  \BibitemOpen
  \bibfield  {author} {\bibinfo {author} {\bibfnamefont {J.~E.}\ \bibnamefont
  {{Chambers}}},\ }\href {\doibase 10.1046/j.1365-8711.1999.02379.x} {\bibfield
   {journal} {\bibinfo  {journal} {\mnras}\ }\textbf {\bibinfo {volume}
  {304}},\ \bibinfo {pages} {793} (\bibinfo {year} {1999})}\BibitemShut
  {NoStop}%
\bibitem [{\citenamefont {{Leinhardt}}\ and\ \citenamefont
  {{Stewart}}(2012)}]{leiste12}%
  \BibitemOpen
  \bibfield  {author} {\bibinfo {author} {\bibfnamefont {Z.~M.}\ \bibnamefont
  {{Leinhardt}}}\ and\ \bibinfo {author} {\bibfnamefont {S.~T.}\ \bibnamefont
  {{Stewart}}},\ }\href {\doibase 10.1088/0004-637X/745/1/79} {\bibfield
  {journal} {\bibinfo  {journal} {\apj}\ }\textbf {\bibinfo {volume} {745}},\
  \bibinfo {eid} {79} (\bibinfo {year} {2012})},\ \Eprint
  {http://arxiv.org/abs/1106.6084} {arXiv:1106.6084 [astro-ph.EP]} \BibitemShut
  {NoStop}%
\bibitem [{\citenamefont {{Genda}}\ \emph
  {et~al.}(2012{\natexlab{a}})\citenamefont {{Genda}}, \citenamefont
  {{Kokubo}},\ and\ \citenamefont {{Ida}}}]{2012ApJ...744..137G}%
  \BibitemOpen
  \bibfield  {author} {\bibinfo {author} {\bibfnamefont {H.}~\bibnamefont
  {{Genda}}}, \bibinfo {author} {\bibfnamefont {E.}~\bibnamefont {{Kokubo}}}, \
  and\ \bibinfo {author} {\bibfnamefont {S.}~\bibnamefont {{Ida}}},\ }\href
  {\doibase 10.1088/0004-637X/744/2/137} {\bibfield  {journal} {\bibinfo
  {journal} {\apj}\ }\textbf {\bibinfo {volume} {744}},\ \bibinfo {eid} {137}
  (\bibinfo {year} {2012}{\natexlab{a}})},\ \Eprint
  {http://arxiv.org/abs/1109.4330} {arXiv:1109.4330 [astro-ph.EP]} \BibitemShut
  {NoStop}%
\bibitem [{\citenamefont {{Hanslmeier}}\ and\ \citenamefont
  {{Dvorak}}(1984)}]{1984A&A...132..203H}%
  \BibitemOpen
  \bibfield  {author} {\bibinfo {author} {\bibfnamefont {A.}~\bibnamefont
  {{Hanslmeier}}}\ and\ \bibinfo {author} {\bibfnamefont {R.}~\bibnamefont
  {{Dvorak}}},\ }\href@noop {} {\bibfield  {journal} {\bibinfo  {journal}
  {\aap}\ }\textbf {\bibinfo {volume} {132}},\ \bibinfo {pages} {203} (\bibinfo
  {year} {1984})}\BibitemShut {NoStop}%
\bibitem [{\citenamefont {{Eggl}}\ and\ \citenamefont
  {{Dvorak}}(2010)}]{2010LNP...790..431E}%
  \BibitemOpen
  \bibfield  {author} {\bibinfo {author} {\bibfnamefont {S.}~\bibnamefont
  {{Eggl}}}\ and\ \bibinfo {author} {\bibfnamefont {R.}~\bibnamefont
  {{Dvorak}}},\ }in\ \href {\doibase 10.1007/978-3-642-04458-8_9} {\emph
  {\bibinfo {booktitle} {Lecture Notes in Physics, Berlin Springer Verlag}}},\
  \bibinfo {series} {Lecture Notes in Physics, Berlin Springer Verlag}, Vol.\
  \bibinfo {volume} {790},\ \bibinfo {editor} {edited by\ \bibinfo {editor}
  {\bibfnamefont {J.}~\bibnamefont {{Souchay}}}\ and\ \bibinfo {editor}
  {\bibfnamefont {R.}~\bibnamefont {{Dvorak}}}}\ (\bibinfo {year} {2010})\ pp.\
  \bibinfo {pages} {431--480}\BibitemShut {NoStop}%
\bibitem [{\citenamefont {{Dvorak}}\ \emph {et~al.}(2012)\citenamefont
  {{Dvorak}}, \citenamefont {{Eggl}}, \citenamefont {{S{\"u}li}}, \citenamefont
  {{S{\'a}ndor}}, \citenamefont {{Galiazzo}},\ and\ \citenamefont
  {{Pilat-Lohinger}}}]{dvoegg12}%
  \BibitemOpen
  \bibfield  {author} {\bibinfo {author} {\bibfnamefont {R.}~\bibnamefont
  {{Dvorak}}}, \bibinfo {author} {\bibfnamefont {S.}~\bibnamefont {{Eggl}}},
  \bibinfo {author} {\bibfnamefont {{\'A}.}~\bibnamefont {{S{\"u}li}}},
  \bibinfo {author} {\bibfnamefont {Z.}~\bibnamefont {{S{\'a}ndor}}}, \bibinfo
  {author} {\bibfnamefont {M.}~\bibnamefont {{Galiazzo}}}, \ and\ \bibinfo
  {author} {\bibfnamefont {E.}~\bibnamefont {{Pilat-Lohinger}}},\ }in\ \href
  {\doibase 10.1063/1.4745576} {\emph {\bibinfo {booktitle} {American Institute
  of Physics Conference Series}}},\ \bibinfo {series} {American Institute of
  Physics Conference Series}, Vol.\ \bibinfo {volume} {1468},\ \bibinfo
  {editor} {edited by\ \bibinfo {editor} {\bibfnamefont {M.}~\bibnamefont
  {{Robnik}}}\ and\ \bibinfo {editor} {\bibfnamefont {V.~G.}\ \bibnamefont
  {{Romanovski}}}}\ (\bibinfo {year} {2012})\ pp.\ \bibinfo {pages}
  {137--147}\BibitemShut {NoStop}%
\bibitem [{\citenamefont {{Izidoro}}\ \emph {et~al.}(2013)\citenamefont
  {{Izidoro}}, \citenamefont {{de Souza Torres}}, \citenamefont {{Winter}},\
  and\ \citenamefont {{Haghighipour}}}]{izides13}%
  \BibitemOpen
  \bibfield  {author} {\bibinfo {author} {\bibfnamefont {A.}~\bibnamefont
  {{Izidoro}}}, \bibinfo {author} {\bibfnamefont {K.}~\bibnamefont {{de Souza
  Torres}}}, \bibinfo {author} {\bibfnamefont {O.~C.}\ \bibnamefont
  {{Winter}}}, \ and\ \bibinfo {author} {\bibfnamefont {N.}~\bibnamefont
  {{Haghighipour}}},\ }\href {\doibase 10.1088/0004-637X/767/1/54} {\bibfield
  {journal} {\bibinfo  {journal} {\apj}\ }\textbf {\bibinfo {volume} {767}},\
  \bibinfo {eid} {54} (\bibinfo {year} {2013})},\ \Eprint
  {http://arxiv.org/abs/1302.1233} {arXiv:1302.1233 [astro-ph.EP]} \BibitemShut
  {NoStop}%
\bibitem [{\citenamefont {{Lunine}}\ \emph {et~al.}(2011)\citenamefont
  {{Lunine}}, \citenamefont {{O'brien}}, \citenamefont {{Raymond}},
  \citenamefont {{Morbidelli}}, \citenamefont {{Quinn}},\ and\ \citenamefont
  {{Graps}}}]{lunobr11}%
  \BibitemOpen
  \bibfield  {author} {\bibinfo {author} {\bibfnamefont {J.~I.}\ \bibnamefont
  {{Lunine}}}, \bibinfo {author} {\bibfnamefont {D.~P.}\ \bibnamefont
  {{O'brien}}}, \bibinfo {author} {\bibfnamefont {S.~N.}\ \bibnamefont
  {{Raymond}}}, \bibinfo {author} {\bibfnamefont {A.}~\bibnamefont
  {{Morbidelli}}}, \bibinfo {author} {\bibfnamefont {T.}~\bibnamefont
  {{Quinn}}}, \ and\ \bibinfo {author} {\bibfnamefont {A.~L.}\ \bibnamefont
  {{Graps}}},\ }\href {\doibase 10.1166/asl.2011.1212} {\bibfield  {journal}
  {\bibinfo  {journal} {Advanced Science Letters}\ }\textbf {\bibinfo {volume}
  {4}},\ \bibinfo {pages} {325} (\bibinfo {year} {2011})}\BibitemShut {NoStop}%
\bibitem [{\citenamefont {{Alexander}}\ and\ \citenamefont
  {{Agnor}}(1998)}]{aleagn98}%
  \BibitemOpen
  \bibfield  {author} {\bibinfo {author} {\bibfnamefont {S.~G.}\ \bibnamefont
  {{Alexander}}}\ and\ \bibinfo {author} {\bibfnamefont {C.~B.}\ \bibnamefont
  {{Agnor}}},\ }\href {\doibase 10.1006/icar.1998.5905} {\bibfield  {journal}
  {\bibinfo  {journal} {\icarus}\ }\textbf {\bibinfo {volume} {132}},\ \bibinfo
  {pages} {113} (\bibinfo {year} {1998})}\BibitemShut {NoStop}%
\bibitem [{\citenamefont {{Agnor}}\ \emph {et~al.}(1999)\citenamefont
  {{Agnor}}, \citenamefont {{Canup}},\ and\ \citenamefont
  {{Levison}}}]{agncan99}%
  \BibitemOpen
  \bibfield  {author} {\bibinfo {author} {\bibfnamefont {C.~B.}\ \bibnamefont
  {{Agnor}}}, \bibinfo {author} {\bibfnamefont {R.~M.}\ \bibnamefont
  {{Canup}}}, \ and\ \bibinfo {author} {\bibfnamefont {H.~F.}\ \bibnamefont
  {{Levison}}},\ }\href {\doibase 10.1006/icar.1999.6201} {\bibfield  {journal}
  {\bibinfo  {journal} {\icarus}\ }\textbf {\bibinfo {volume} {142}},\ \bibinfo
  {pages} {219} (\bibinfo {year} {1999})}\BibitemShut {NoStop}%
\bibitem [{\citenamefont {{Asphaug}}(2010)}]{asp10}%
  \BibitemOpen
  \bibfield  {author} {\bibinfo {author} {\bibfnamefont {E.}~\bibnamefont
  {{Asphaug}}},\ }\href {\doibase 10.1016/j.chemer.2010.01.004} {\bibfield
  {journal} {\bibinfo  {journal} {Chemie der Erde / Geochemistry}\ }\textbf
  {\bibinfo {volume} {70}},\ \bibinfo {pages} {199} (\bibinfo {year}
  {2010})}\BibitemShut {NoStop}%
\bibitem [{\citenamefont {{Agnor}}\ and\ \citenamefont
  {{Asphaug}}(2004{\natexlab{b}})}]{agnasp04}%
  \BibitemOpen
  \bibfield  {author} {\bibinfo {author} {\bibfnamefont {C.}~\bibnamefont
  {{Agnor}}}\ and\ \bibinfo {author} {\bibfnamefont {E.}~\bibnamefont
  {{Asphaug}}},\ }\href {\doibase 10.1086/425158} {\bibfield  {journal}
  {\bibinfo  {journal} {\apjl}\ }\textbf {\bibinfo {volume} {613}},\ \bibinfo
  {pages} {L157} (\bibinfo {year} {2004}{\natexlab{b}})}\BibitemShut {NoStop}%
\bibitem [{\citenamefont {{Marcus}}\ \emph {et~al.}(2009)\citenamefont
  {{Marcus}}, \citenamefont {{Stewart}}, \citenamefont {{Sasselov}},\ and\
  \citenamefont {{Hernquist}}}]{marste09}%
  \BibitemOpen
  \bibfield  {author} {\bibinfo {author} {\bibfnamefont {R.~A.}\ \bibnamefont
  {{Marcus}}}, \bibinfo {author} {\bibfnamefont {S.~T.}\ \bibnamefont
  {{Stewart}}}, \bibinfo {author} {\bibfnamefont {D.}~\bibnamefont
  {{Sasselov}}}, \ and\ \bibinfo {author} {\bibfnamefont {L.}~\bibnamefont
  {{Hernquist}}},\ }\href {\doibase 10.1088/0004-637X/700/2/L118} {\bibfield
  {journal} {\bibinfo  {journal} {\apjl}\ }\textbf {\bibinfo {volume} {700}},\
  \bibinfo {pages} {L118} (\bibinfo {year} {2009})},\ \Eprint
  {http://arxiv.org/abs/0907.0234} {arXiv:0907.0234 [astro-ph.EP]} \BibitemShut
  {NoStop}%
\bibitem [{\citenamefont {{Kokubo}}\ and\ \citenamefont
  {{Genda}}(2010)}]{kokgen10}%
  \BibitemOpen
  \bibfield  {author} {\bibinfo {author} {\bibfnamefont {E.}~\bibnamefont
  {{Kokubo}}}\ and\ \bibinfo {author} {\bibfnamefont {H.}~\bibnamefont
  {{Genda}}},\ }\href {\doibase 10.1088/2041-8205/714/1/L21} {\bibfield
  {journal} {\bibinfo  {journal} {\apjl}\ }\textbf {\bibinfo {volume} {714}},\
  \bibinfo {pages} {L21} (\bibinfo {year} {2010})},\ \Eprint
  {http://arxiv.org/abs/1003.4384} {arXiv:1003.4384 [astro-ph.EP]} \BibitemShut
  {NoStop}%
\bibitem [{\citenamefont {{Genda}}\ \emph
  {et~al.}(2012{\natexlab{b}})\citenamefont {{Genda}}, \citenamefont
  {{Kokubo}},\ and\ \citenamefont {{Ida}}}]{genkok12}%
  \BibitemOpen
  \bibfield  {author} {\bibinfo {author} {\bibfnamefont {H.}~\bibnamefont
  {{Genda}}}, \bibinfo {author} {\bibfnamefont {E.}~\bibnamefont {{Kokubo}}}, \
  and\ \bibinfo {author} {\bibfnamefont {S.}~\bibnamefont {{Ida}}},\ }\href
  {\doibase 10.1088/0004-637X/744/2/137} {\bibfield  {journal} {\bibinfo
  {journal} {\apj}\ }\textbf {\bibinfo {volume} {744}},\ \bibinfo {eid} {137}
  (\bibinfo {year} {2012}{\natexlab{b}})},\ \Eprint
  {http://arxiv.org/abs/1109.4330} {arXiv:1109.4330 [astro-ph.EP]} \BibitemShut
  {NoStop}%
\bibitem [{\citenamefont {{Canup}}\ \emph {et~al.}(2013)\citenamefont
  {{Canup}}, \citenamefont {{Barr}},\ and\ \citenamefont
  {{Crawford}}}]{canbar13}%
  \BibitemOpen
  \bibfield  {author} {\bibinfo {author} {\bibfnamefont {R.~M.}\ \bibnamefont
  {{Canup}}}, \bibinfo {author} {\bibfnamefont {A.~C.}\ \bibnamefont {{Barr}}},
  \ and\ \bibinfo {author} {\bibfnamefont {D.~A.}\ \bibnamefont {{Crawford}}},\
  }\href {\doibase 10.1016/j.icarus.2012.10.011} {\bibfield  {journal}
  {\bibinfo  {journal} {\icarus}\ }\textbf {\bibinfo {volume} {222}},\ \bibinfo
  {pages} {200} (\bibinfo {year} {2013})}\BibitemShut {NoStop}%
\bibitem [{\citenamefont {{Melosh}}\ and\ \citenamefont
  {{Ryan}}(1997)}]{melrya97}%
  \BibitemOpen
  \bibfield  {author} {\bibinfo {author} {\bibfnamefont {H.~J.}\ \bibnamefont
  {{Melosh}}}\ and\ \bibinfo {author} {\bibfnamefont {E.~V.}\ \bibnamefont
  {{Ryan}}},\ }\href {\doibase 10.1006/icar.1997.5797} {\bibfield  {journal}
  {\bibinfo  {journal} {\icarus}\ }\textbf {\bibinfo {volume} {129}},\ \bibinfo
  {pages} {562} (\bibinfo {year} {1997})}\BibitemShut {NoStop}%
\bibitem [{\citenamefont {{Maindl}}\ \emph
  {et~al.}(2014{\natexlab{a}})\citenamefont {{Maindl}}, \citenamefont
  {{Dvorak}}, \citenamefont {{Speith}},\ and\ \citenamefont
  {{Sch{\"a}fer}}}]{maidvo13b}%
  \BibitemOpen
  \bibfield  {author} {\bibinfo {author} {\bibfnamefont {T.~I.}\ \bibnamefont
  {{Maindl}}}, \bibinfo {author} {\bibfnamefont {R.}~\bibnamefont {{Dvorak}}},
  \bibinfo {author} {\bibfnamefont {R.}~\bibnamefont {{Speith}}}, \ and\
  \bibinfo {author} {\bibfnamefont {C.}~\bibnamefont {{Sch{\"a}fer}}},\
  }\href@noop {} {\bibfield  {journal} {\bibinfo  {journal} {ArXiv e-print
  arXiv:1401.0045}\ } (\bibinfo {year} {2014}{\natexlab{a}})},\ \Eprint
  {http://arxiv.org/abs/1401.0045} {arXiv:1401.0045 [astro-ph.EP]} \BibitemShut
  {NoStop}%
\bibitem [{\citenamefont {{Tillotson}}(1962)}]{til62}%
  \BibitemOpen
  \bibfield  {author} {\bibinfo {author} {\bibfnamefont {J.~H.}\ \bibnamefont
  {{Tillotson}}},\ }\href@noop {} {\emph {\bibinfo {title} {Metallic Equations
  of State for Hypervelocity Impact}}},\ \bibinfo {type} {Tech. Rep.}\ \bibinfo
  {number} {General Atomic Report GA-3216}\ (\bibinfo  {institution} {General
  Dynamics},\ \bibinfo {address} {San Diego, CA},\ \bibinfo {year}
  {1962})\BibitemShut {NoStop}%
\bibitem [{\citenamefont {{Melosh}}(1989)}]{mel89}%
  \BibitemOpen
  \bibfield  {author} {\bibinfo {author} {\bibfnamefont {H.~J.}\ \bibnamefont
  {{Melosh}}},\ }\href@noop {} {\emph {\bibinfo {title} {Research supported by
  NASA.~New York, Oxford University Press (Oxford Monographs on Geology and
  Geophysics, No.~11), 1989, 253 p.}}}\ (\bibinfo  {publisher} {Oxford
  University Press, New York},\ \bibinfo {year} {1989})\BibitemShut {NoStop}%
\bibitem [{\citenamefont {{Sch{\"a}fer}}\ \emph {et~al.}(2007)\citenamefont
  {{Sch{\"a}fer}}, \citenamefont {{Speith}},\ and\ \citenamefont
  {{Kley}}}]{schspe07}%
  \BibitemOpen
  \bibfield  {author} {\bibinfo {author} {\bibfnamefont {C.}~\bibnamefont
  {{Sch{\"a}fer}}}, \bibinfo {author} {\bibfnamefont {R.}~\bibnamefont
  {{Speith}}}, \ and\ \bibinfo {author} {\bibfnamefont {W.}~\bibnamefont
  {{Kley}}},\ }\href {\doibase 10.1051/0004-6361:20077354} {\bibfield
  {journal} {\bibinfo  {journal} {\aap}\ }\textbf {\bibinfo {volume} {470}},\
  \bibinfo {pages} {733} (\bibinfo {year} {2007})},\ \Eprint
  {http://arxiv.org/abs/0705.2672} {arXiv:0705.2672} \BibitemShut {NoStop}%
\bibitem [{\citenamefont {{Benz}}\ and\ \citenamefont
  {{Asphaug}}(1994)}]{benasp94}%
  \BibitemOpen
  \bibfield  {author} {\bibinfo {author} {\bibfnamefont {W.}~\bibnamefont
  {{Benz}}}\ and\ \bibinfo {author} {\bibfnamefont {E.}~\bibnamefont
  {{Asphaug}}},\ }\href {\doibase 10.1006/icar.1994.1009} {\bibfield  {journal}
  {\bibinfo  {journal} {\icarus}\ }\textbf {\bibinfo {volume} {107}},\ \bibinfo
  {pages} {98} (\bibinfo {year} {1994})}\BibitemShut {NoStop}%
\bibitem [{\citenamefont {{Grady}}\ and\ \citenamefont
  {{Kipp}}(1980)}]{grakip80}%
  \BibitemOpen
  \bibfield  {author} {\bibinfo {author} {\bibfnamefont {D.~E.}\ \bibnamefont
  {{Grady}}}\ and\ \bibinfo {author} {\bibfnamefont {M.~E.}\ \bibnamefont
  {{Kipp}}},\ }\href {\doibase
  {http://dx.doi.org/10.1016/0148-9062(80)91361-3}} {\bibfield  {journal}
  {\bibinfo  {journal} {{International Journal of Rock Mechanics and Mining
  Sciences \& Geomechanics Abstracts}}\ }\textbf {\bibinfo {volume} {17}},\
  \bibinfo {pages} {147} (\bibinfo {year} {1980})}\BibitemShut {NoStop}%
\bibitem [{\citenamefont {{Weibull}}(1939)}]{wei39}%
  \BibitemOpen
  \bibfield  {author} {\bibinfo {author} {\bibfnamefont {W.~A.}\ \bibnamefont
  {{Weibull}}},\ }\href@noop {} {\bibfield  {journal} {\bibinfo  {journal}
  {{Ingvetensk. Akad. Handl.}}\ }\textbf {\bibinfo {volume} {151}},\ \bibinfo
  {pages} {5} (\bibinfo {year} {1939})}\BibitemShut {NoStop}%
\bibitem [{\citenamefont {{Benz}}\ and\ \citenamefont
  {{Asphaug}}(1999)}]{benasp99}%
  \BibitemOpen
  \bibfield  {author} {\bibinfo {author} {\bibfnamefont {W.}~\bibnamefont
  {{Benz}}}\ and\ \bibinfo {author} {\bibfnamefont {E.}~\bibnamefont
  {{Asphaug}}},\ }\href {\doibase 10.1006/icar.1999.6204} {\bibfield  {journal}
  {\bibinfo  {journal} {\icarus}\ }\textbf {\bibinfo {volume} {142}},\ \bibinfo
  {pages} {5} (\bibinfo {year} {1999})},\ \Eprint
  {http://arxiv.org/abs/arXiv:astro-ph/9907117} {arXiv:astro-ph/9907117}
  \BibitemShut {NoStop}%
\bibitem [{\citenamefont {{Nakamura}}\ \emph {et~al.}(2007)\citenamefont
  {{Nakamura}}, \citenamefont {{Michel}},\ and\ \citenamefont
  {{Setoh}}}]{nakmic07}%
  \BibitemOpen
  \bibfield  {author} {\bibinfo {author} {\bibfnamefont {A.~M.}\ \bibnamefont
  {{Nakamura}}}, \bibinfo {author} {\bibfnamefont {P.}~\bibnamefont
  {{Michel}}}, \ and\ \bibinfo {author} {\bibfnamefont {M.}~\bibnamefont
  {{Setoh}}},\ }\href {\doibase 10.1029/2006JE002757} {\bibfield  {journal}
  {\bibinfo  {journal} {Journal of Geophysical Research (Planets)}\ }\textbf
  {\bibinfo {volume} {112}},\ \bibinfo {eid} {E02001} (\bibinfo {year}
  {2007})}\BibitemShut {NoStop}%
\bibitem [{\citenamefont {{Lange}}\ \emph {et~al.}(1984)\citenamefont
  {{Lange}}, \citenamefont {{Ahrens}},\ and\ \citenamefont
  {{Boslough}}}]{lanahr84}%
  \BibitemOpen
  \bibfield  {author} {\bibinfo {author} {\bibfnamefont {M.~A.}\ \bibnamefont
  {{Lange}}}, \bibinfo {author} {\bibfnamefont {T.~J.}\ \bibnamefont
  {{Ahrens}}}, \ and\ \bibinfo {author} {\bibfnamefont {M.~B.}\ \bibnamefont
  {{Boslough}}},\ }\href {\doibase 10.1016/0019-1035(84)90084-8} {\bibfield
  {journal} {\bibinfo  {journal} {\icarus}\ }\textbf {\bibinfo {volume} {58}},\
  \bibinfo {pages} {383} (\bibinfo {year} {1984})}\BibitemShut {NoStop}%
\bibitem [{\citenamefont {{Sch{\"a}fer}}(2005)}]{sch05}%
  \BibitemOpen
  \bibfield  {author} {\bibinfo {author} {\bibfnamefont {C.}~\bibnamefont
  {{Sch{\"a}fer}}},\ }\emph {\bibinfo {title} {Application of Smooth Particle
  Hydrodynamics to selected aspects of planet formation}},\ \href@noop {}
  {\bibinfo {type} {Dissertation}},\ \bibinfo  {school}
  {Eberhard-Karls-Universit{\"a}t T{\"u}bingen} (\bibinfo {year}
  {2005})\BibitemShut {NoStop}%
\bibitem [{\citenamefont {{Maindl}}\ \emph {et~al.}(2013)\citenamefont
  {{Maindl}}, \citenamefont {{Sch{\"a}fer}}, \citenamefont {{Speith}},
  \citenamefont {{S{\"u}li}}, \citenamefont {{Forg{\'a}cs-Dajka}},\ and\
  \citenamefont {{Dvorak}}}]{maisch13}%
  \BibitemOpen
  \bibfield  {author} {\bibinfo {author} {\bibfnamefont {T.~I.}\ \bibnamefont
  {{Maindl}}}, \bibinfo {author} {\bibfnamefont {C.}~\bibnamefont
  {{Sch{\"a}fer}}}, \bibinfo {author} {\bibfnamefont {R.}~\bibnamefont
  {{Speith}}}, \bibinfo {author} {\bibfnamefont {{\'A}.}~\bibnamefont
  {{S{\"u}li}}}, \bibinfo {author} {\bibfnamefont {E.}~\bibnamefont
  {{Forg{\'a}cs-Dajka}}}, \ and\ \bibinfo {author} {\bibfnamefont
  {R.}~\bibnamefont {{Dvorak}}},\ }\href {\doibase 10.1002/asna.201311979}
  {\bibfield  {journal} {\bibinfo  {journal} {Astronomische Nachrichten}\
  }\textbf {\bibinfo {volume} {334}},\ \bibinfo {pages} {996} (\bibinfo {year}
  {2013})}\BibitemShut {NoStop}%
\bibitem [{\citenamefont {{Maindl}}\ \emph {et~al.}(2015)\citenamefont
  {{Maindl}}, \citenamefont {{Dvorak}}, \citenamefont {{Lammer}}, \citenamefont
  {{G{\"u}del}}, \citenamefont {{Sch{\"a}fer}}, \citenamefont {{Speith}},
  \citenamefont {{Odert}}, \citenamefont {{Erkaev}}, \citenamefont
  {{Kislyakova}},\ and\ \citenamefont {{Pilat-Lohinger}}}]{maidvo14}%
  \BibitemOpen
  \bibfield  {author} {\bibinfo {author} {\bibfnamefont {T.~I.}\ \bibnamefont
  {{Maindl}}}, \bibinfo {author} {\bibfnamefont {R.}~\bibnamefont {{Dvorak}}},
  \bibinfo {author} {\bibfnamefont {H.}~\bibnamefont {{Lammer}}}, \bibinfo
  {author} {\bibfnamefont {M.}~\bibnamefont {{G{\"u}del}}}, \bibinfo {author}
  {\bibfnamefont {C.}~\bibnamefont {{Sch{\"a}fer}}}, \bibinfo {author}
  {\bibfnamefont {R.}~\bibnamefont {{Speith}}}, \bibinfo {author}
  {\bibfnamefont {P.}~\bibnamefont {{Odert}}}, \bibinfo {author} {\bibfnamefont
  {N.~V.}\ \bibnamefont {{Erkaev}}}, \bibinfo {author} {\bibfnamefont {K.~G.}\
  \bibnamefont {{Kislyakova}}}, \ and\ \bibinfo {author} {\bibfnamefont
  {E.}~\bibnamefont {{Pilat-Lohinger}}},\ }\href {\doibase
  10.1051/0004-6361/201424256} {\bibfield  {journal} {\bibinfo  {journal}
  {\aap}\ }\textbf {\bibinfo {volume} {574}},\ \bibinfo {eid} {A22} (\bibinfo
  {year} {2015})},\ \Eprint {http://arxiv.org/abs/1405.5913} {arXiv:1405.5913
  [astro-ph.EP]} \BibitemShut {NoStop}%
\bibitem [{\citenamefont {{Maindl}}\ and\ \citenamefont
  {{Haghighipour}}(2014)}]{maihag14}%
  \BibitemOpen
  \bibfield  {author} {\bibinfo {author} {\bibfnamefont {T.~I.}\ \bibnamefont
  {{Maindl}}}\ and\ \bibinfo {author} {\bibfnamefont {N.}~\bibnamefont
  {{Haghighipour}}},\ }in\ \href@noop {} {\emph {\bibinfo {booktitle}
  {AAS/Division for Planetary Sciences Meeting Abstracts}}},\ \bibinfo {series}
  {AAS/Division for Planetary Sciences Meeting Abstracts}, Vol.~\bibinfo
  {volume} {46}\ (\bibinfo {year} {2014})\ p.\ \bibinfo {pages}
  {\#415.17}\BibitemShut {NoStop}%
\bibitem [{\citenamefont {{Maindl}}\ and\ \citenamefont
  {{Dvorak}}(2014)}]{maidvo13}%
  \BibitemOpen
  \bibfield  {author} {\bibinfo {author} {\bibfnamefont {T.~I.}\ \bibnamefont
  {{Maindl}}}\ and\ \bibinfo {author} {\bibfnamefont {R.}~\bibnamefont
  {{Dvorak}}},\ }in\ \href {\doibase 10.1017/S1743921313008971} {\emph
  {\bibinfo {booktitle} {IAU Symposium}}},\ \bibinfo {series} {IAU Symposium},
  Vol.\ \bibinfo {volume} {299},\ \bibinfo {editor} {edited by\ \bibinfo
  {editor} {\bibfnamefont {M.}~\bibnamefont {{Booth}}}, \bibinfo {editor}
  {\bibfnamefont {B.~C.}\ \bibnamefont {{Matthews}}}, \ and\ \bibinfo {editor}
  {\bibfnamefont {J.~R.}\ \bibnamefont {{Graham}}}}\ (\bibinfo {year} {2014})\
  pp.\ \bibinfo {pages} {370--373},\ \Eprint {http://arxiv.org/abs/1307.1643}
  {arXiv:1307.1643 [astro-ph.EP]} \BibitemShut {NoStop}%
\bibitem [{\citenamefont {{Maindl}}\ \emph
  {et~al.}(2014{\natexlab{b}})\citenamefont {{Maindl}}, \citenamefont
  {{Dvorak}}, \citenamefont {{Schäfer}},\ and\ \citenamefont
  {{Speith}}}]{maidvo14b}%
  \BibitemOpen
  \bibfield  {author} {\bibinfo {author} {\bibfnamefont {T.~I.}\ \bibnamefont
  {{Maindl}}}, \bibinfo {author} {\bibfnamefont {R.}~\bibnamefont {{Dvorak}}},
  \bibinfo {author} {\bibfnamefont {C.}~\bibnamefont {{Schäfer}}}, \ and\
  \bibinfo {author} {\bibfnamefont {R.}~\bibnamefont {{Speith}}},\ }in\ \href
  {\doibase 10.1017/S1743921314008059} {\emph {\bibinfo {booktitle} {Complex
  Planetary Systems}}},\ \bibinfo {series} {Proceedings of the International
  Astronomical Union}, Vol.~\bibinfo {volume} {9}\ (\bibinfo {year} {2014})\
  pp.\ \bibinfo {pages} {138--141}\BibitemShut {NoStop}%
\bibitem [{\citenamefont {{Haghighipour}}\ and\ \citenamefont
  {{Raymond}}(2007)}]{hagray07}%
  \BibitemOpen
  \bibfield  {author} {\bibinfo {author} {\bibfnamefont {N.}~\bibnamefont
  {{Haghighipour}}}\ and\ \bibinfo {author} {\bibfnamefont {S.~N.}\
  \bibnamefont {{Raymond}}},\ }\href {\doibase 10.1086/520501} {\bibfield
  {journal} {\bibinfo  {journal} {\apj}\ }\textbf {\bibinfo {volume} {666}},\
  \bibinfo {pages} {436} (\bibinfo {year} {2007})},\ \Eprint
  {http://arxiv.org/abs/astro-ph/0702706} {astro-ph/0702706} \BibitemShut
  {NoStop}%
\bibitem [{\citenamefont {{Wolszczan}}\ and\ \citenamefont
  {{Frail}}(1992)}]{1992Natur.355..145W}%
  \BibitemOpen
  \bibfield  {author} {\bibinfo {author} {\bibfnamefont {A.}~\bibnamefont
  {{Wolszczan}}}\ and\ \bibinfo {author} {\bibfnamefont {D.~A.}\ \bibnamefont
  {{Frail}}},\ }\href {\doibase 10.1038/355145a0} {\bibfield  {journal}
  {\bibinfo  {journal} {\nat}\ }\textbf {\bibinfo {volume} {355}},\ \bibinfo
  {pages} {145} (\bibinfo {year} {1992})}\BibitemShut {NoStop}%
\bibitem [{\citenamefont {{Mayor}}\ and\ \citenamefont
  {{Queloz}}(1995)}]{1995Natur.378..355M}%
  \BibitemOpen
  \bibfield  {author} {\bibinfo {author} {\bibfnamefont {M.}~\bibnamefont
  {{Mayor}}}\ and\ \bibinfo {author} {\bibfnamefont {D.}~\bibnamefont
  {{Queloz}}},\ }\href {\doibase 10.1038/378355a0} {\bibfield  {journal}
  {\bibinfo  {journal} {\nat}\ }\textbf {\bibinfo {volume} {378}},\ \bibinfo
  {pages} {355} (\bibinfo {year} {1995})}\BibitemShut {NoStop}%
\bibitem [{Note1()}]{Note1}%
  \BibitemOpen
  \bibinfo {note} {BwGRiD (http://www.bw-grid.de), member of the German D-Grid
  initiative, funded by the Ministry for Education and Research
  (Bundesministerium fuer Bildung und Forschung) and the Ministry for Science,
  Research and Arts Baden-Wuerttemberg (Ministerium fuer Wissenschaft,
  Forschung und Kunst Baden-Wuerttemberg).}\BibitemShut {Stop}%
\end{thebibliography}%

\end{document}